\documentclass[12pt]{article}
\usepackage{a4wide}
\usepackage{latexsym}
\usepackage{amsmath}
\usepackage{amsfonts}
\usepackage{cite}
\usepackage{graphicx}
\usepackage{axodraw}

\usepackage{pslatex}
\usepackage[T1]{fontenc}
\usepackage[latin1]{inputenc}

\def\bq{\begin{eqnarray}}
\def\eq{\end{eqnarray}}
\def\eps{\varepsilon}

\newcommand{\mathcalI}{\mbox{\fontencoding{OMS}\fontfamily{cmr}\fontseries{m}\fontshape{it}\fontsize{12pt}{12pt}\selectfont I}}

\begin{document}

\thispagestyle{empty}

\begin{flushright}
  MZ-TH/10-05 \\
  TTK-10-18
\end{flushright}

\vspace{1.5cm}

\begin{center}
  {\Large\bf Feynman graph polynomials\\
  }
  \vspace{1cm}
  {\large Christian Bogner$^1$ and Stefan Weinzierl$^2$\\
  \vspace{6mm}
      {\small \em $^1$ Institut f\"ur Theoretische Teilchenphysik und Kosmologie, RWTH Aachen,}\\
      {\small \em   D - 52056 Aachen, Germany}\\
  \vspace{2mm}
      {\small \em $^2$ Institut f{\"u}r Physik, Universit{\"a}t Mainz,}\\
      {\small \em D - 55099 Mainz, Germany}\\
  } 
\end{center}

\vspace{2cm}

\begin{abstract}\noindent
  {
The integrand of any multi-loop integral is characterised after Feynman parametrisation 
by two polynomials.
In this review we summarise the properties of these polynomials.
Topics covered in this article include among others: 
Spanning trees and spanning forests, the all-minors matrix-tree theorem, 
recursion relations due to contraction and deletion of edges, Dodgson's identity
and matroids.
   }
\end{abstract}

\vspace*{\fill}

\newpage

\section{Introduction}
\label{sect_introduction}

In this review we discuss Feynman graph polynomials.
Let us first motivate the interest in these polynomials. 
The Feynman graph polynomials are of interest from a phenomenological point of view
as well as from a more mathematical perspective.
We start with the phenomenological aspects: 
For the practitioner of perturbative loop calculations the integrand of any multi-loop integral
is characterised after Feynman parametrisation by two polynomials.
These two polynomials, which are called the first and the second Symanzik polynomial, can be read off
directly from the Feynman graph and are the subject of this review.
They have many special properties and these properties can be used to derive algorithms for
the computation of loop integrals.

In recent years the graph polynomials have received additional attention from a more formal point of 
view \cite{Bloch:2005,Bloch:2008jk,Bloch:2007,Bloch:2008,Andre:2008aa,Brown:2008,Brown:2009a,Brown:2009b,Schnetz:2008mp,Schnetz:2009,Marcolli:2008cy,Aluffi:2008sy,Aluffi:2008rw,Aluffi:2009b,Aluffi:2009a,Marcolli_book,Bergbauer:2009yu,Laporta:2002pg,Laporta:2004rb,Laporta:2008sx,Bailey:2008ib,Blumlein:2009,Belkale:2003,Belkale:2003b,Bierenbaum:2003ud,Bogner:2007cr,Bogner:2007mn,Grozin:2008tp,Patterson:PhD}.
Feynman integrals are now considered as non-trivial examples of mixed Hodge structures and motives.
The zero sets of the graph polynomials play a crucial role in this setting.

Graph polynomials have a long history dating back to Kirchhoff \cite{Kirchhoff:1847}. 
There are well-established books on this subject \cite{Eden,Hwa,Nakanishi,Todorov,Zavialov,Smirnov:2006ry}. 
However the field has evolved, new insights have been added and old results have been re-discovered.
As a consequence the available information is scattered over the literature.
What is missing is a concise summary of the properties of these polynomials.
With this review we hope to fill this gap.
We have tried to make this article accessible both to the phenomenological oriented physicist interested
in loop calculations as well as to the mathematician interested in the properties of Feynman integrals.
In addition we include in a few places results which are new and cannot be found in the literature.

This review is organised as follows: 
In the next section we recall some basic facts about multi-loop Feynman integrals.
We introduce the two polynomials associated to a Feynman graph 
and give a first method for their computation.
Sect.~\ref{sect_singularities} is devoted to the singularities of a Feynman integral.
There are two aspects to it.
If we fix the external momenta and take them to lie in the Euclidean region
the singularities of the Feynman integral after integration are of ultraviolet or infrared
origin. They arise -- apart from a possible overall ultraviolet divergence -- from the regions
in Feynman parameter space where one of the two graph polynomials vanishes.
For the second aspect we give up the restriction on the Euclidean region and view the Feynman integral
as a function of the external momenta. As we vary the external momenta, additional threshold
singularities may arise. Necessary conditions for them are given by Landau's equations.
In sect.~\ref{sect_spanning_trees} we start to introduce concepts of graph theory
and define for a graph its spanning trees and spanning forests.
These concepts lead to a second method for the computation of the graph polynomials, such that the
two graph polynomials can be directly read off from the topology of the graph.
Sect.~\ref{sect_matrix_tree_theorem} introduces the Laplacian of a graph and states 
the matrix-tree theorem.
This in turn provides a third method for the computation of the two graph polynomials.
This method is well suited for computer algebra, as it involves just the computation
of a determinant of a matrix.
The matrix is easily constructed from the data defining the graph.
In sect.~\ref{sect_deletion_and_contraction} the two operations of deleting and contracting an edge
are studied in detail. This leads to a fourth and recursive method 
for the computation of the two graph polynomials.
In addition we discuss in this section the multivariate Tutte polynomial and Dodgson's identity.
Sect.~\ref{sect_duality} is devoted to the dual of a (planar) graph.
The Kirchhoff polynomial and the first Symanzik polynomial exchange their role when going from a
graph to its dual.
Sect.~\ref{sect_matroids} is of a more formal character. We introduce matroids, which provide
a natural generalisation of graphs.
Within matroid theory, some things are simpler: For example there is always the dual of a matroid,
whereas for graphs we were restricted to planar graphs.
Matroid theory provides in addition an answer to the question under which conditions two topologically
different graphs have the same Kirchhoff polynomial.
Finally, sect.~\ref{sect_conclusions} contains our conclusions.

\section{Feynman integrals}
\label{sect_feynman_integrals}

\begin{figure}
\begin{center}
\includegraphics[bb= 115 610 300 725]{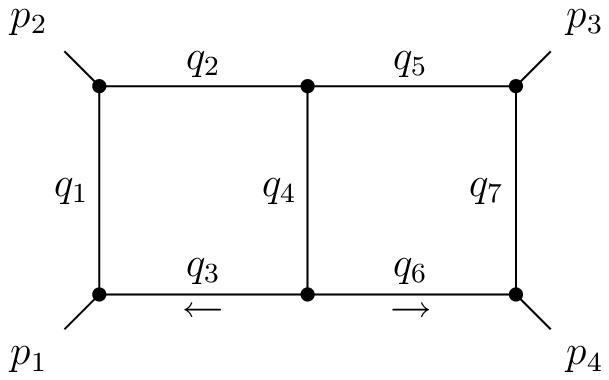}
\end{center}
\caption{\label{fig1} The ``double box''-graph:
A two-loop Feynman diagram with four external lines and seven internal lines.
The momenta flowing out along the external lines are labelled $p_1$, ..., $p_4$, 
the momenta flowing through the internal lines are labelled $q_1$, ..., $q_7$.}
\end{figure}

In this section we recall some basic facts about multi-loop Feynman integrals.
We introduce the two polynomials associated to a Feynman graph and give a first method for their computation.
We will work in a space-time of $D$ dimensions.
To set the scene let us consider a scalar Feynman graph $G$ with $m$ external lines and $n$ internal lines.
Fig.~\ref{fig1} shows an example.
In this example there are four external lines and seven internal lines. 
The momenta flowing in or out through the external lines
are labelled $p_1$, $p_2$, $p_3$ and $p_4$ and can be taken as fixed $D$-dimensional vectors.
They are constrained by momentum conservation: If all momenta are taken to flow outwards,
momentum conservation requires that
\bq
 p_1 + p_2 + p_3 + p_4 & = & 0.
\eq
At each vertex of a graph we have again momentum conservation: The sum of all momenta flowing into the vertex
equals the sum of all momenta flowing out of the vertex.

A graph, where the external momenta determine uniquely all internal momenta is called a tree graph.
It can be shown that such a graph does not contain any closed circuit.
In contrast, graphs which do contain one or more closed circuits are called loop graphs.
If we have to specify besides the external momenta in addition 
$l$ internal momenta in order to determine uniquely all
internal momenta we say that the graph contains $l$ loops.
In this sense, a tree graph is a graph with zero loops and the graph in fig.~\ref{fig1} contains two loops.
In more mathematical terms the number $l$ is known as the cyclomatic number or the first Betti number of the graph.
Let us agree that we label the $l$ additional internal momenta by $k_1$ to $k_l$.
In the example of fig.~\ref{fig1} there are two independent loop momenta. 
We can choose them to be $k_1=q_3$ and $k_2=q_6$.
Then all other internal momenta are expressed in terms of $k_1$, $k_2$ and the external momenta
$p_1$, ..., $p_4$:
\bq
\begin{array}{lll}
 q_1 = k_1 - p_1,
&
 q_2 = k_1 - p_1 - p_2,
&
 q_4 = k_1 + k_2,
 \\
 q_5 = k_2 - p_3 - p_4,
 &
 q_7 = k_2 - p_4.
 & \\
\end{array}
\eq
In general, each momentum flowing through an internal line is given
as a linear combination of the external 
momenta $p$ and the loop momenta $k$ with coefficients $-1$, $0$ or $1$:
\bq
\label{eq_internal_mom}
 q_i & = & \sum\limits_{j=1}^l \lambda_{ij} k_j + \sum\limits_{j=1}^m \sigma_{ij} p_j,
 \;\;\; \lambda_{ij}, \sigma_{ij} \in \{-1,0,1\}.
\eq
We associate to a Feynman graph the Feynman integral 
\bq
\label{eq_basic_feynman_integral}
I_G  & = &
 \left( \mu^2 \right)^{\nu-l D/2}
 \int \prod\limits_{r=1}^{l} \frac{d^Dk_r}{i\pi^{\frac{D}{2}}}\;
 \prod\limits_{j=1}^{n} \frac{1}{(-q_j^2+m_j^2)^{\nu_j}},
 \;\;\;\;\;\;
 \nu = \sum_{j=1}^n \nu_j.
\eq
This Feynman integral depends on the Feynman graph $G$ and the external momenta $p$. 
The graph $G$ together with the independent loop momenta $k$ and the external momenta $p$
fixes all internal momenta according to eq.~(\ref{eq_internal_mom}).
In addition $I_G$ depends on
the masses $m_j$ of the particles corresponding to the internal line $j$, as well as
integer numbers $\nu_j$, specifying the power to which each propagator is raised.
The parameter $\mu$ is an arbitrary mass scale. We have multiplied the integral
with a factor $(\mu^2)^{\nu-l D/2}$, this ensures that $I_G$ is dimensionless.

How to perform the integration over the loop momenta?
The first step is to convert the products of propagators into a sum.
This can be done with the Feynman parameter technique.
In its full generality it is also applicable to cases, where each factor in the denominator is raised to 
some power $\nu_j$.
The formula reads:
\bq
 \prod\limits_{j=1}^{n} \frac{1}{P_{j}^{\nu_{j}}} 
 & = &
 \frac{\Gamma(\nu)}{\prod\limits_{j=1}^{n} \Gamma(\nu_{j})} \;\;
 \int\limits_{x_j \ge 0} d^nx \; \delta(1-\sum\limits_{j=1}^{n} x_{j})
 \frac{\left( \prod\limits_{j=1}^{n} x_{j}^{\nu_{j}-1} \right)}
      {\left( \sum\limits_{j=1}^{n} x_{j} P_{j} \right)^{\nu}},
 \;\;\;\;\;\;
 \nu = \sum\limits_{j=1}^{n} \nu_{j}. 
\eq
We use this formula with $P_j=-q_j^2+m_j^2$.
Applied to eq.~(\ref{eq_basic_feynman_integral}) we have
\bq
 \sum\limits_{i=1}^{n} x_{i} P_{i} & = & \sum\limits_{i=1}^{n} x_{i} (-q_i^2+m_i^2).
\eq
Now one can use translational invariance of the $D$-dimensional loop integrals and shift each loop
momentum $k_r$ to complete the square, such that the integrand depends only on $k_r^2$.
Then all $D$-dimensional loop integrals can be performed.
As the integrals over the Feynman parameters still remain,
this allows us to treat the
$D$-dimensional loop integrals for Feynman parameter integrals.
One arrives at the following Feynman parameter integral \cite{Itzykson:1980rh}:
\bq
\label{eq_feynman_parameter_integral}
I_G  & = &
 \frac{\Gamma(\nu-lD/2)}{\prod\limits_{j=1}^{n}\Gamma(\nu_j)}
 \int\limits_{x_j \ge 0} d^nx \; \delta(1-\sum_{i=1}^n x_i)
 \left( \prod\limits_{j=1}^{n}\,dx_j\,x_j^{\nu_j-1} \right)\,\frac{{\mathcal U}^{\nu-(l+1) D/2}}
 {{\mathcal F}^{\nu-l D/2}}.
\eq
The functions ${\mathcal U}$ and $\mathcal F$ depend on the Feynman parameters $x_j$.
If one expresses
\bq
\label{eq_poly_calc_1}
 \sum\limits_{j=1}^{n} x_{j} (-q_j^2+m_j^2)
 & = & 
 - \sum\limits_{r=1}^{l} \sum\limits_{s=1}^{l} k_r M_{rs} k_s + \sum\limits_{r=1}^{l} 2 k_r \cdot Q_r + J,
\eq
where $M$ is a $l \times l$ matrix with scalar entries and $Q$ is a $l$-vector
with four-vectors as entries,
one obtains
\bq
\label{eq_poly_calc_2}
 {\mathcal U} = \mbox{det}(M),
 & &
 {\mathcal F} = \mbox{det}(M) \left( J + Q M^{-1} Q \right)/\mu^2.
\eq
The functions ${\mathcal U}$ and ${\mathcal F}$ are called graph polynomials and are the subject of this review.
They are polynomials in the Feynman parameters and -- as we will show later -- can be derived from the topology
of the underlying graph.
The polynomials ${\mathcal U}$ and ${\mathcal F}$ have the following properties:
\begin{itemize}
\item They are homogeneous in the Feynman parameters, 
${\mathcal U}$ is of degree $l$, ${\mathcal F}$ is of degree $l+1$.
\item ${\mathcal U}$ is linear in each Feynman parameter. If all internal masses are zero, then
also ${\mathcal F}$ is linear in each Feynman parameter.
\item In expanded form each monomial of ${\mathcal U}$ has coefficient $+1$.
\end{itemize}
We call ${\mathcal U}$ the first Symanzik polynomial and ${\mathcal F}$ the second Symanzik polynomial.
Eqs.~(\ref{eq_poly_calc_1}) and (\ref{eq_poly_calc_2}) allow us to calculate these polynomials for a given graph.
We will learn several alternative ways to determine these polynomials later, 
but for the moment it is instructive to go through this exercise
for the graph of fig.~\ref{fig1}.
We will consider the case
\bq
\label{specification_double_box}
 & & p_1^2 = 0, \;\;\; p_2^2 = 0, \;\;\; p_3^2 = 0, \;\;\; p_4^2 = 0,
 \nonumber \\
 & & m_1 = m_2 = m_3 = m_4 = m_5 = m_6 = m_7 = 0.
\eq
We define
\bq
 s = \left(p_1+p_2\right)^2=\left(p_3+p_4\right)^2,
 & &
 t = \left(p_2+p_3\right)^2=\left(p_1+p_4\right)^2.
\eq
We have
\bq
 \sum\limits_{j=1}^7 x_j \left(-q_j^2\right) & = &
 - \left(x_1+x_2+x_3+x_4\right) k_1^2 - 2 x_4 k_1 \cdot k_2 - \left( x_4+x_5+x_6+x_7\right) k_2^2
 \\
 & &
 + 2 \left[ x_1 p_1 + x_2 \left( p_1 + p_2 \right) \right] \cdot k_1
 + 2 \left[ x_5 \left( p_3 + p_4 \right) + x_7 p_4 \right] \cdot k_2
 - \left( x_2 + x_5 \right) s.
 \nonumber
\eq
In comparing with eq.~(\ref{eq_poly_calc_1})
we find
\bq
 M & = & \left( \begin{array}{cc}
 x_1+x_2+x_3+x_4 & x_4 \\
 x_4 & x_4+x_5+x_6+x_7 \\
 \end{array} \right),
 \nonumber \\
 Q & = & \left( \begin{array}{c}
          x_1 p_1 + x_2 \left( p_1 + p_2 \right) \\
          x_5 \left( p_3 + p_4 \right) + x_7 p_4 \\
          \end{array} \right),
 \nonumber \\
 J & = & \left( x_2 + x_5 \right) \left(-s\right).
\eq
Plugging this into eq.~(\ref{eq_poly_calc_2})
we obtain the graph polynomials as
\bq
\label{result_U_and_F_double_box}
{\mathcal U} & = & \left( x_1+x_2+x_3 \right) \left( x_5+x_6+x_7 \right) + x_4 \left( x_1+x_2+x_3+x_5+x_6+x_7 \right),
 \nonumber \\
{\mathcal F} & = & \left[ x_2 x_3 \left( x_4+x_5+x_6+x_7 \right)
                        + x_5 x_6 \left( x_1+x_2+x_3+x_4 \right)
                        + x_2 x_4 x_6 + x_3 x_4 x_5 \right] \left( \frac{-s}{\mu^2} \right)
 \nonumber \\
 & &
      + x_1 x_4 x_7 \left( \frac{-t}{\mu^2} \right).
\eq
We see in this example that ${\mathcal U}$ is of degree $2$ and ${\mathcal F}$ is of degree $3$.
Each polynomial is linear in each Feynman parameter. Furthermore, when we write ${\mathcal U}$
in expanded form
\bq 
 {\mathcal U} & = &
  x_1 x_5 + x_1 x_6 + x_1 x_7
+ x_2 x_5 + x_2 x_6 + x_2 x_7
+ x_3 x_5 + x_3 x_6 + x_3 x_7
 \nonumber \\
 & &
+ x_1 x_4 + x_2 x_4 + x_3 x_4 + x_4 x_5 + x_4 x_6 + x_4 x_7,
\eq
each term has coefficient $+1$.


\section{Singularities}
\label{sect_singularities}

In this section we briefly discuss singularities of Feynman integrals.
There are two aspects to it. 
First we fix the external momenta and take them to lie in the Euclidean region.
We may encounter singularities in the Feynman integral. 
These singularities are of ultraviolet or infrared origin and require regularisation.
We briefly discuss how they are related to the vanishing of the two graph polynomials
${\mathcal U}$ and ${\mathcal F}$.
For the second aspect we consider the Feynman integrals as a function of the external momenta,
which are now allowed to lie in the physical region. 
Landau's equations give a necessary condition for a singularity to occur in the Feynman integral as
we vary the external momenta.

It often occurs that the Feynman integral as given in eq.~(\ref{eq_basic_feynman_integral})
or in eq.~(\ref{eq_feynman_parameter_integral}) is an ill-defined and divergent expression
when considered in $D=4$ dimensions.
These divergences are related to ultraviolet or infrared singularities.
Dimensional regularisation is usually employed to regulate these divergences.
Within dimensional regularisation one considers the Feynman integral in $D=4-2\eps$ dimensions.
Going away from the integer value $D=4$ regularises the integral.
In $D=4-2\eps$ dimensions the Feynman integral has a Laurent expansion in the parameter $\eps$.
The poles of the Laurent series correspond to the original divergences of the integral in
four dimensions.

From the Feynman parameter integral in eq.~(\ref{eq_feynman_parameter_integral}) we see that
there are three possibilities how poles in $\eps$ can arise:
First of all the Gamma-function $\Gamma(\nu-l D/2)$ of the prefactor can give rise to 
a (single) pole if the argument of this function is close to zero or to a negative integer
value. This divergence is called the overall ultraviolet divergence.

Secondly, we consider the polynomial ${\mathcal U}$. 
Depending on the exponent $\nu-(l+1)D/2$ of ${\mathcal U}$ 
the vanishing of the polynomial ${\mathcal U}$ 
in some part of the integration region can lead to poles in $\eps$ after integration.
As mentioned in the previous section, each term of the expanded form of the polynomial ${\mathcal U}$ 
has coefficient $+1$, therefore ${\mathcal U}$ can only vanish if some of the Feynman
parameters are equal to zero.
In other words, ${\mathcal U}$ is non-zero (and positive) inside the integration region, but
may vanish on the boundary of the integration region.
Poles in $\eps$ resulting from the vanishing of ${\mathcal U}$ are related to
ultraviolet sub-divergences.

Thirdly, we consider the polynomial ${\mathcal F}$. 
In an analytic calculation one often considers the Feynman integral in the Euclidean region.
The Euclidean region is defined as the region, where all invariants
$(p_{i_1}+p_{i_2}+...+p_{i_k})^2$ are negative or zero, and all internal masses are positive or 
zero.
The result in the physical region is then obtained by analytic continuation.
It can be shown that in the Euclidean region the polynomial ${\mathcal F}$ 
is also non-zero (and positive) inside the integration region.
Therefore under the assumption that the external kinematics is within the Euclidean region
the polynomial ${\mathcal F}$ can only vanish on the boundary of the integration region, 
similar to what has been observed for the the polynomial ${\mathcal U}$.
Depending on the exponent $\nu-l D/2$ of ${\mathcal F}$
the vanishing of the polynomial ${\mathcal F}$ on the boundary of
the integration region may lead to poles in $\eps$ after integration.
These poles are related to infrared divergences.

The Feynman integral $I_G$ as given in eq.~(\ref{eq_feynman_parameter_integral})
depends through the polynomial ${\mathcal F}$ on the external momenta $p_j$.
We can also discuss $I_G$ as a function of the $p_j$'s without restricting the external
kinematics to the Euclidean region.
Doing so, the region where the polynomial ${\mathcal F}$ vanishes is no longer restricted
to the boundary of the Feynman parameter integration region
and we may encounter zeros of the polynomial
${\mathcal F}$ inside the integration region.
The vanishing of ${\mathcal F}$ may in turn result in divergences after integration.
These singularities are called Landau singularities.
Necessary conditions for the occurrence of a Landau singularity are given as follows:
A Landau singularity may occur if ${\mathcal F}=0$ and if there exists a subset $S$ of $\{1,...,n\}$
such that
\bq
\label{Landau_condition}
 x_i & = & 0
 \;\;\; \mbox{for} \; i \in S
 \nonumber \\
 \mbox{and} \;\;\;\;
 \frac{\partial}{\partial x_j} {\mathcal F} & = & 0
 \;\;\; \mbox{for} \; j \in \{1,...,n\}\backslash S.
\eq
The case corresponding to $S = \emptyset$ is called the leading Landau singularity, and cases
corresponding to $S \neq \emptyset$ are called non-leading singularities.
It is sufficient to focus on the leading Landau singularity, since a non-leading singularity
is the leading Landau singularity of a sub-graph of $G$ obtained by contracting the propagators
corresponding to the Feynman parameters $x_i$ with $i \in S$.

Let us now consider the leading Landau singularity of a graph $G$ with $m$ external lines.
We view the Feynman integral as a function of the external momenta and 
a solution of the Landau equations is given by a set of external momenta
\bq 
\label{solution_landau_equations}
\{ p_1, p_2, ..., p_m \}
\eq
satisfying momentum conservation and eq.~(\ref{Landau_condition}).
If the momenta in eq.~(\ref{solution_landau_equations}) define a one-dimensional sub-space,
we call the Landau singularity a normal threshold.
In this case all external momenta are collinear.
If on the contrary the momenta in eq.~(\ref{solution_landau_equations}) define a
higher-dimensional sub-space, we speak of an anomalous threshold.

We give a simple example for a Feynman integral with an ultraviolet divergence
and a normal threshold. 
\begin{figure}
\begin{center}
\includegraphics[bb= 125 655 242 720]{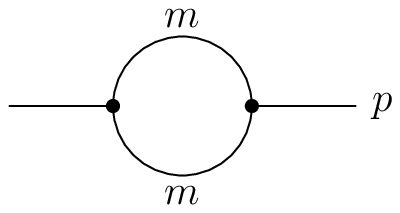}
\end{center}
\caption{\label{fig20} 
The one-loop two-point function with equal masses. This graph shows a normal threshold for $p^2=4m^2$.
}
\end{figure}    
The graph in fig.~\ref{fig20} shows a one-loop two-point
function. The internal propagators correspond to particles with mass $m$.
This graph corresponds in $D=4-2\eps$ dimensions to the Feynman integral
\bq
 I_G & = &
 \left( \mu^2 \right)^{\eps}
 \int \frac{d^Dk}{i\pi^{\frac{D}{2}}}\;
 \frac{1}{(-k^2+m^2)(-(k-p)^2+m^2)}.
\eq
Introducing Feynman parameters one obtains the form of eq.~(\ref{eq_feynman_parameter_integral}):
\bq
 I_G & = & \Gamma(\eps) \int\limits_0^1 dx \; 
  \left[ x(1-x) \left( \frac{-p^2}{\mu^2} \right) + \frac{m^2}{\mu^2} \right]^{-\eps}.
\eq
This integral is easily evaluated with standard techniques \cite{Weinzierl:2006qs}:
\bq 
 I_G
 & = & 
 \frac{1}{\eps} - \gamma_E + 2 - \ln \frac{m^2}{\mu^2}
 + \sqrt{1-\frac{1}{x}} \ln \frac{\sqrt{1-x}-\sqrt{-x}}{\sqrt{1-x}+\sqrt{-x}}
 + {\cal O}(\eps),
\;\;\;\;\; x = \frac{p^2}{4m^2}.
\eq
Here, $\gamma_E$ denotes Euler's constant.
The $1/\eps$-term corresponds to an ultraviolet divergence.
As a function of $p^2$ the integral has a normal threshold at $p^2=4m^2$.
The normal threshold manifests itself as a branch point in the complex $p^2$-plane.


\section{Spanning trees and spanning forests}
\label{sect_spanning_trees}

In this section we start to introduce concepts of graph theory.
We define spanning trees and spanning forests. These concepts lead to a second method for
the computation of the graph polynomials.
We consider a connected graph $G$ with $m$ external lines and $n$ internal lines.
Let $r$ be the number of vertices of the graph $G$.
We denote the set of internal edges of the graph $G$ by
\bq
 \{ e_1, e_2, ..., e_n \}
\eq
and the set of vertices by
\bq
 \{ v_1, v_2, ..., v_r \}.
\eq
As before we denote by $l$ the first Betti number of the graph (or in physics jargon: the number of loops).
We have the relation
\bq
 l & = & n - r + 1.
\eq
If we would allow for disconnected graphs, the corresponding formula for the first Betti number would be
$n-r+k$, where $k$ is the number of connected components.
A spanning tree for the graph $G$ is a sub-graph $T$ of $G$
satisfying the following requirements:
\begin{itemize}
\item $T$ contains all the vertices of $G$,
\item the first Betti number of $T$ is zero,
\item $T$ is connected.
\end{itemize}
If $T$ is a spanning tree for $G$, then it can be obtained from $G$ by deleting $l$ edges.
In general a given graph $G$ has several spanning trees. We will later obtain a formula which counts the number
of spanning trees for a given graph $G$.
A spanning forest for the graph $G$ is a sub-graph $F$ of $G$ satisfying just the first two requirements:
\begin{itemize}
\item $F$ contains all the vertices of $G$,
\item the first Betti number of $F$ is zero.
\end{itemize}
It is not required that a spanning forest is connected.
If $F$ has $k$ connected components, we say that $F$ is a $k$-forest.
A spanning tree is a spanning $1$-forest.
If $F$ is a spanning $k$-forest for $G$, then it can be obtained from $G$ by deleting $l+k-1$ edges.
\begin{figure}
\begin{center}
\includegraphics[bb= 130 620 520 720,width=0.7\textwidth]{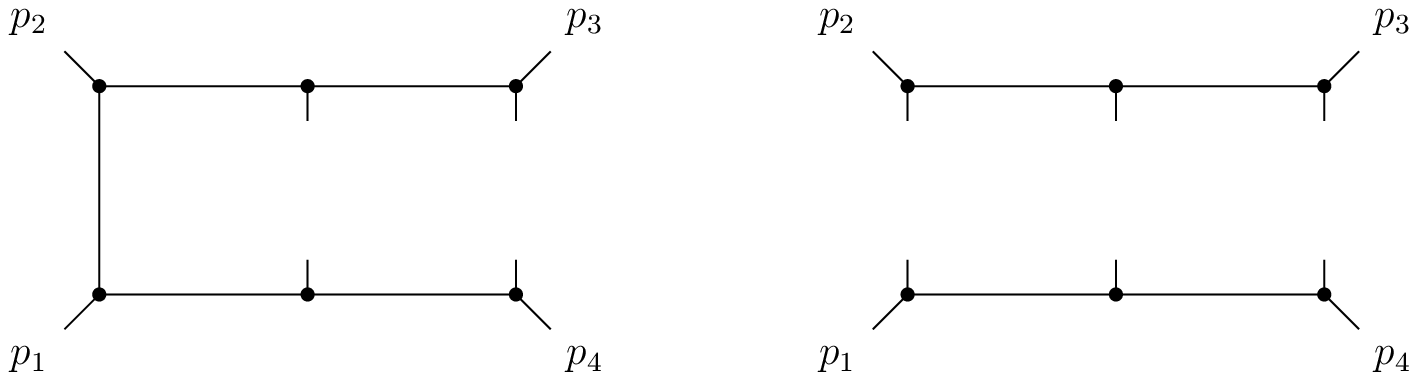}
\end{center}
\caption{\label{fig2} The left picture shows a spanning tree for the graph of fig.~\ref{fig1},
the right picture shows a spanning $2$-forest for the same graph.
The spanning tree is obtained by deleting edges $4$ and $7$, the spanning $2$-forest is obtained
by deleting edges $1$, $4$ and $7$.
}
\end{figure}    
Fig.~\ref{fig2} shows an example for a spanning tree and a spanning $2$-forest for the graph of fig.~\ref{fig1}.

We denote by ${\cal T}$ the set of spanning forests of $G$ and by ${\cal T}_k$ the set of spanning $k$-forests of $G$.
Obviously, we can write ${\cal T}$ as the disjoint union
\bq
 {\cal T} & = & \bigcup\limits_{k=1}^{r} {\cal T}_k.
\eq
${\cal T}_1$ is the set of spanning trees.
For an element of ${\cal T}_k$ we write
\bq
 \left( T_1, T_2, ..., T_k \right) & \in & {\cal T}_k.
\eq
The $T_i$ are the connected components of the $k$-forest. They are necessarily trees.
We denote by $P_{T_i}$ the set of external momenta attached to $T_i$.
For the example of the $2$-forest in the right picture of fig.~\ref{fig2} we have
\bq
 P_{T_1} = \{ p_2, p_3 \},
 & &
 P_{T_2} = \{ p_1, p_4 \}.
\eq
The spanning trees and the spanning $2$-forests of a graph $G$ are closely related to the 
graph polynomials ${\mathcal U}$ and ${\mathcal F}$ of the graph.
We have
\bq
\label{eq_poly_calc_3}	
 {\mathcal U}
 & = & 
 \sum\limits_{T\in {\mathcal T}_1} \;
     \prod\limits_{e_i\notin T} x_i,
 \nonumber\\
 {\mathcal F}
 & = & 
 \sum\limits_{(T_1,T_2)\in {\mathcal T}_2} \;
     \left( \prod\limits_{e_i\notin (T_1,T_2)} x_i \right) 
     \left( \sum\limits_{p_j\in P_{T_1}} \sum\limits_{p_k\in P_{T_2}} \frac{p_j \cdot p_k}{\mu^2} \right)
 \; + \; {\mathcal U} \sum\limits_{i=1}^n x_i \frac{m_i^2}{\mu^2}.
\eq
The sum is over all spanning trees for ${\mathcal U}$, and over all spanning $2$-forests
in the first term of the formula for ${\mathcal F}$.
Eq.~(\ref{eq_poly_calc_3}) provides a second method for the computation of the graph polynomials
${\mathcal U}$ and ${\mathcal F}$.
Let us first look at the formula for ${\mathcal U}$. For each spanning tree $T$ we take the edges $e_i$,
which have been removed from the graph $G$ to obtain $T$. The product of the
corresponding Feynman parameters $x_i$ gives a monomial.
The first formula says, that ${\mathcal U}$ is the sum of all the monomials obtained from all spanning trees.
The formula for ${\mathcal F}$ has two parts: One part is related to the external momenta and the other part
involves the masses.
The latter is rather simple and we write
\bq
\label{def_F_0}
 {\mathcal F} & = & {\mathcal F}_0 + {\mathcal U} \sum\limits_{i=1}^n x_i \frac{m_i^2}{\mu^2}.
\eq
We focus on the polynomial ${\mathcal F}_0$. 
Here the $2$-forests are relevant. For each $2$-forest $(T_1,T_2)$
we consider again the edges $e_i$,
which have been removed from the graph $G$ to obtain $(T_1,T_2)$. 
The product of the corresponding Feynman parameters $x_i$ defines again a monomial,
which in addition is multiplied by a quantity which depends on the external momenta.
We define the square of the sum of momenta through the cut lines of $(T_1,T_2)$ by
\bq
 s_{(T_1,T_2)} & = & \left( \sum\limits_{e_j\notin (T_1,T_2)} q_j \right)^2.
\eq
Here we assumed for simplicity that the orientation of the momenta of the cut internal lines are chosen
such that all cut momenta flow from $T_1$ to $T_2$ (or alternatively that all cut momenta flow from $T_2$
to $T_1$, but not mixed).
From momentum conservation it follows that
the sum of the momenta flowing through the cut lines out of $T_1$
is equal to the negative of the sum of the external momenta of $T_1$.
With the same reasoning 
the sum of the momenta flowing through the cut lines into $T_2$
is equal to the sum of the external momenta of $T_2$.
Therefore we can equally write
\bq
 s_{(T_1,T_2)} & = &
   - \left( \sum\limits_{p_i\in P_{T_1}} p_i \right) \cdot \left( \sum\limits_{p_j\in P_{T_2}} p_j \right)
\eq
and ${\mathcal F}_0$ is given by
\bq
 {\mathcal F}_0
 & = & 
 \sum\limits_{(T_1,T_2)\in {\mathcal T}_2} \;
     \left( \prod\limits_{e_i\notin (T_1,T_2)} x_i \right)
     \left( \frac{-s_{(T_1,T_2)}}{\mu^2} \right).
\eq
Since we have to remove $l$ edges from $G$ to obtain a spanning tree and $(l+1)$ edges to obtain
a spanning $2$-forest, it follows that ${\mathcal U}$ and ${\mathcal F}$ are homogeneous in the Feynman
parameters of degree $l$ and $(l+1)$, respectively.
From the fact, that an internal edge can be removed at most once, it follows that ${\mathcal U}$ and
${\mathcal F}_0$ are linear in each Feynman parameter.
Finally it is obvious from eq.~(\ref{eq_poly_calc_3}) that each monomial in the expanded form of ${\mathcal U}$
has coefficient $+1$.

\begin{figure}
\begin{center}
\includegraphics[bb= 125 630 250 705]{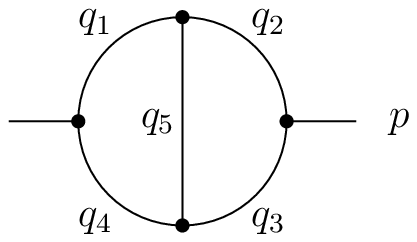}
\end{center}
\caption{\label{fig3} A two-loop two-point graph.}
\end{figure}

\begin{figure}
\begin{center}
\includegraphics[bb= 125 585 440 715]{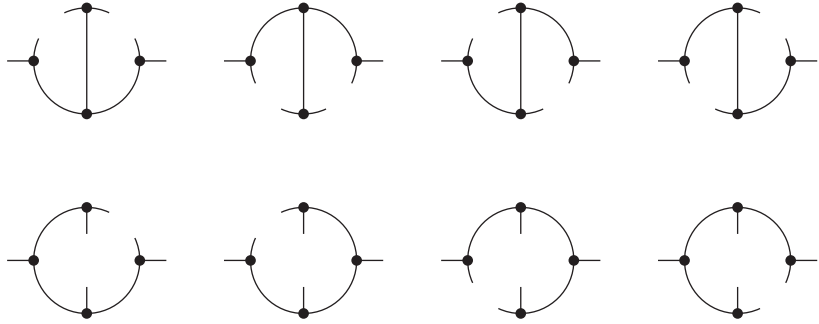}
\end{center}
\caption{\label{fig4} The set of spanning trees for the two-loop two-point graph of fig.~\ref{fig3}.}
\end{figure}

Let us look at an example. Fig.~\ref{fig3} shows the graph of a two-loop two-point integral.
We take again all internal masses to be zero.
The set of all spanning trees for this graph is shown in fig.~\ref{fig4}.
There are eight spanning trees.
\begin{figure}
\begin{center}
\includegraphics[bb= 125 585 510 715]{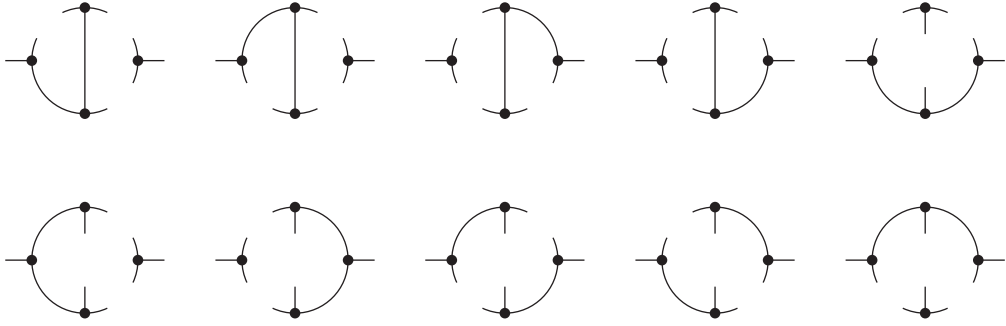}
\end{center}
\caption{\label{fig5} The set of spanning $2$-forests for the two-loop two-point graph of fig.~\ref{fig3}.}
\end{figure}    
Fig.~\ref{fig5} shows the set of spanning $2$-forests for this graph. There are ten spanning $2$-forests.
The last example in each row of fig.~\ref{fig5} does not contribute to the graph polynomial ${\mathcal F}$, since
the momentum sum flowing through all cut lines is zero. Therefore we have in this case $s_{(T_1,T_2)}=0$.
In all other cases we have $s_{(T_1,T_2)}=p^2$. We arrive therefore at the graph polynomials
\bq
\label{graph_polynomials_ex_tbubble}
 {\mathcal U} & = & (x_1+x_4)(x_2+x_3) + (x_1+x_2+x_3+x_4)x_5,
 \nonumber \\
 {\mathcal F} & = & 
      \left[ (x_1+x_2)(x_3+x_4)x_5
      +x_1x_4(x_2+x_3)
      +x_2x_3(x_1+x_4) \right]
      \left( \frac{-p^2}{\mu^2} \right).
\eq


\section{The matrix-tree theorem}
\label{sect_matrix_tree_theorem}

In this section we introduce the Laplacian of a graph. 
The Laplacian is a matrix constructed from the topology of the graph.
The determinant of a minor of this matrix where the $i$-th row and column have been deleted
gives us the Kirchhoff polynomial of the graph, which in turn
upon a simple substitution leads to the first Symanzik polynomial.
We then show how this construction generalises for the second Symanzik polynomial.
This provides a third method for the computation of the two graph polynomials.
This method is very well suited for computer algebra systems, as it involves just the computation
of a determinant of a matrix. The matrix is easily constructed from the data defining the graph.

We begin with the Kirchhoff polynomial of a graph. 
This polynomial is defined by
\bq
 {\mathcal K}\left(x_1,...,x_n\right)
 & = & 
 \sum\limits_{T\in {\mathcal T}_1} \;
     \prod\limits_{e_j \in T} x_j.
\eq
The definition is very similar to the expression for the first Symanzik polynomial in eq.~(\ref{eq_poly_calc_3}).
Again we have a sum over all spanning trees, but this time we take for each spanning tree the monomial of the Feynman 
parameters corresponding to the edges which have not been removed.
The Kirchhoff polynomial is therefore homogeneous of degree $(n-l)$ in the Feynman parameters.
There is a simple relation between the Kirchhoff polynomial ${\mathcal K}$ 
and the first Symanzik polynomial ${\mathcal U}$:
\bq
\label{convert_U_K}
 {\mathcal U}(x_1,...,x_n) = x_1 ... x_n \; {\mathcal K}\left(\frac{1}{x_1},...,\frac{1}{x_n}\right),
 & &
 {\mathcal K}(x_1,...,x_n) = x_1 ... x_n \; {\mathcal U}\left(\frac{1}{x_1},...,\frac{1}{x_n}\right).
\eq
These equations are immediately evident from the fact that ${\mathcal U}$ and ${\mathcal K}$ are
homogeneous polynomials which are linear in each variable together with
the fact that a monomial corresponding to a 
specific spanning tree in one polynomial contains exactly those Feynman parameters which are not 
in the corresponding monomial in the other polynomial.

We now define the Laplacian of a graph $G$ with $n$ edges and $r$ vertices as a
$r \times r$-matrix $L$, whose entries are given by \cite{Tutte:1984,Stanley:1998}
\bq
 L_{ij} & = & \left\{ \begin{array}{rl} 
                      \sum x_k & \mbox{if $i =j$ and edge $e_k$ is attached to $v_i$ and is not a self-loop,} \\
                      - \sum x_k & \mbox{if $i \neq j$ and edge $e_k$ connects $v_i$ and $v_j$.}\\
                      \end{array} \right.
\eq
The graph may contain multiple edges and self-loops.
\begin{figure}
\begin{center}
\includegraphics[bb= 125 660 330 725]{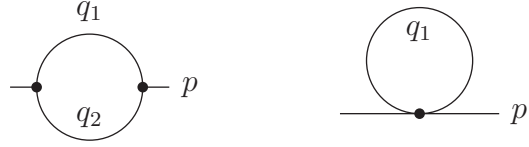}
\end{center}
\caption{\label{fig6} The left picture shows a graph with a double edge, the right picture shows a graph with a self-loop.}
\end{figure}    
We speak of a multiple edge, if two vertices are connected by more than one edge.
We speak of a self-loop if an edge starts and ends at the same vertex. In the physics literature a self-loop is
known as a tadpole. Fig.~\ref{fig6} shows a simple example for a double edge and a self-loop.
If the vertices $v_i$ and $v_j$ are connected by two edges $e_{k_1}$ and $e_{k_2}$, then the Laplacian depends only on the
sum $x_{k_1}+x_{k_2}$.
If an edge $e_k$ is a self-loop attached to a vertex $v_i$, then it does not contribute to the Laplacian.

Let us consider an example: The Laplacian of the two-loop two-point graph of fig.~\ref{fig3} is given by
\bq
\label{example_laplacian}
 L & = & \left( \begin{array}{cccc}
                x_1+x_4 & -x_1 & 0 & -x_4 \\ 
                -x_1 & x_1+x_2+x_5 & -x_2 & -x_5 \\ 
                0 & -x_2 & x_2+x_3 & -x_3 \\ 
                -x_4 & -x_5 & -x_3 & x_3+x_4+x_5 \\ 
                \end{array} \right).
\eq
In the sequel we will need minors of the matrix $L$ and it is convenient to introduce the
following notation:
For a $r \times r$ matrix $A$ we denote by $A[i_1,...,i_k;j_1,...,j_k]$ the
$(r-k)\times(r-k)$ matrix, which is obtained from $A$ by deleting the rows $i_1$, ..., $i_k$ and the 
columns $j_1$, ..., $j_k$.
For $A[i_1,...,i_k;i_1,...,i_k]$ we will simply write $A[i_1,...,i_k]$.

Let $v_i$ be an arbitrary vertex of $G$. The matrix-tree theorem states \cite{Tutte:1984}
\bq
\label{matrix_tree_theorem}
 {\mathcal K} & = & \det \; L[i],
\eq
i.e. the Kirchhoff polynomial is given by the determinant of the minor of the Laplacian, where
the $i$-th row and column have been removed.
One can choose for $i$ any number between $1$ and $r$.

Choosing for example $i=4$ in eq.~(\ref{example_laplacian}) one finds
for the Kirchhoff polynomial of the two-loop two-point graph of fig.~\ref{fig3}
\bq
 {\mathcal K} & = &
         \left| \begin{array}{ccc}
                x_1+x_4 & -x_1 & 0 \\ 
                -x_1 & x_1+x_2+x_5 & -x_2 \\ 
                0 & -x_2 & x_2+x_3 \\ 
                \end{array} \right|
 \nonumber \\
 & = &
  x_1 x_2 ( x_3 + x_4 ) + ( x_1 + x_2 ) x_3 x_4 
+ \left( x_1 x_2 + x_1 x_3 + x_2 x_4 + x_3 x_4 \right) x_5.
\eq
Using eq.~(\ref{convert_U_K}) one recovers the first Symanzik polynomial of this graph as
given in eq.~(\ref{graph_polynomials_ex_tbubble}).

The matrix-tree theorem allows to determine the number of spanning trees of a given graph
$G$. Setting $x_1=...=x_n=1$, each monomial in ${\mathcal K}$ and ${\mathcal U}$ reduces
to $1$. There is exactly one monomial for each spanning tree, therefore one obtains
\bq
 \left| {\mathcal T}_1 \right| & = &
 {\mathcal K}(1,...,1)
 = {\mathcal U}(1,...,1).
\eq

\begin{figure}
\begin{center}
\includegraphics[bb= 125 650 385 720]{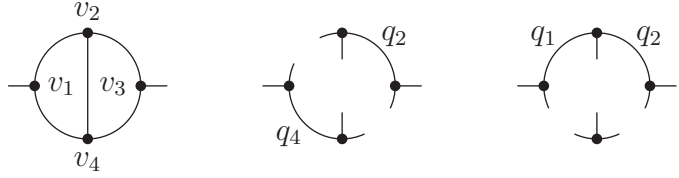}
\end{center}
\caption{\label{fig7} The left picture shows the labelling of the vertices for the two-loop two-point
function. 
The middle and the right picture show the two $2$-forests contributing to ${\mathcal T}_2^{I,J}$
with $I=(2,4)$ and $J=(3,4)$.}
\end{figure}

The matrix-tree theorem as in eq.~(\ref{matrix_tree_theorem}) relates the determinant of 
the minor of the Laplacian, where the
$i$-th row and the $i$-th column have been deleted to a sum over the spanning trees of the graph.
There are two generalisations we can think of:
\begin{enumerate}
\item We delete more than one row and column.
\item We delete different rows and columns, i.e. we delete row $i$ and column $j$ with $i \neq j$.
\end{enumerate}
The all-minors matrix-tree theorem relates the determinant of the corresponding minor to a 
specific sum over spanning forests \cite{Chaiken:1982,Chen:1982,Moon:1994}.
To state this theorem we need some notation:
We consider a graph with $r$ vertices.
Let $I=(i_1,...,i_k)$ with $1\le i_1 < ... < i_k \le r$ denote the rows, which we delete from the
Laplacian, and let $J=(j_1,...,j_k)$ with $1\le j_1 < ... < j_k \le r$ denote the columns to be deleted
from the Laplacian.
We set $|I|=i_1+...+i_k$ and $|J|=j_1+...+j_k$.
We denote by ${\mathcal T}_k^{I,J}$ the spanning $k$-forests, such that 
each tree of an element of ${\mathcal T}_k^{I,J}$ contains exactly one vertex $v_{i_\alpha}$ and
exactly one vertex $v_{j_\beta}$.
The set ${\mathcal T}_k^{I,J}$ is a sub-set of all spanning $k$-forests.
We now consider an element $F$ of ${\mathcal T}_k^{I,J}$. 
Since the element $F$ is a $k$-forest, it consists
therefore of
$k$ trees and we can write it as
\bq 
 F = \left( T_1, ..., T_k \right) & \in & {\mathcal T}_k^{I,J}.
\eq
We can label the trees such that $v_{i_1} \in T_1$, ..., $v_{i_k} \in T_k$.
By assumption, each tree $T_\alpha$ contains also exactly one vertex from the set
$\{v_{j_1},...,v_{j_k}\}$, although not necessarily in the order $v_{j_\alpha} \in T_\alpha$.
In general it will be in a different order, which we can specify by a permutation
$\pi_F \in S_k$:
\bq
 v_{j_\alpha} \in T_{\pi_F(\alpha)}.
\eq
The all-minors matrix-tree theorem reads then
\bq
\label{all_minors_tree_theorem}
 \det \; L[I,J] & = & (-1)^{|I|+|J|} \sum\limits_{F\in {\mathcal T}_k^{I,J}}
 \mbox{sign}(\pi_F) \prod\limits_{e_j \in F} x_j.
\eq
In the special case $I=J$ this reduces to
\bq
 \det \; L[I] & = & \sum\limits_{F\in {\mathcal T}_k^{I,I}} \;
 \prod\limits_{e_j \in F} x_j.
\eq
If we specialise further to $I=J=(i)$, the sum equals the sum over all spanning trees
(since each spanning $1$-forest of ${\mathcal T}_1^{(i),(i)}$ necessarily contains the vertex $v_i$).
We recover the classical matrix-tree theorem:
\bq
 \det \; L[i] & = & \sum\limits_{T\in {\mathcal T}_1} \;
 \prod\limits_{e_j \in T} x_j.
\eq
Let us illustrate the all-minors matrix-tree theorem with an example. We consider again the two-loop two-point graph with the 
labelling of the vertices as shown in fig.~\ref{fig7}.
Taking as an example
\bq
 I = (2,4)
 \;\;\mbox{and}\;\;
 J = (3,4)
\eq
we find for the determinant of $L[I;J]$:
\bq
\label{example_all_minors}
 \det \; L[2,4;3,4] & = & 
 \left| \begin{array}{cc}
 x_1+x_4 & -x_1 \\
 0 & -x_2 \\
 \end{array} \right|
 = -x_1 x_2 - x_2 x_4.
\eq
On the other hand there are exactly two $2$-forests, such that in each $2$-forest the vertices
$v_2$ and $v_3$ are contained in one tree, while the vertex $v_4$ is contained in the other tree.
These two $2$-forests are shown in fig.~\ref{fig7}.
The monomials corresponding to these two $2$-trees are $x_1 x_2$ and $x_2 x_4$, respectively.
The permutation $\pi_F$ is in both cases the identity and with $|I|=6$, $|J|=7$ we have
an overall minus sign
\bq
 (-1)^{|I|+|J|} & = & -1.
\eq
Therefore, the right hand side of eq.~(\ref{all_minors_tree_theorem}) equals $-x_1x_2 - x_2 x_4$,
showing the agreement with the result of eq.~(\ref{example_all_minors}).

Eq.~(\ref{matrix_tree_theorem}) together with eq.~(\ref{convert_U_K}) allows to determine 
the first Symanzik polynomial ${\mathcal U}$ from the Laplacian of the graph.
We may ask if also the polynomial ${\mathcal F}_0$ can be obtained in a similar way.
We consider again a graph $G$ with $n$ internal edges $(e_1,...,e_n)$, 
$r$ internal vertices $(v_1,...,v_r)$ and $m$ external legs.
We attach $m$ additional vertices $v_{r+1}$, ..., $v_{r+m}$ to the ends of the external legs and view
the former external legs as additional edges $e_{n+1}$, ..., $e_{n+m}$.
This defines a new graph $\tilde{G}$.
As before we associate the parameters $x_i$ to the edges $e_i$ ($1\le i \le n$) 
and new parameters $z_j$ to the edges $e_{n+j}$ ($1\le j \le m$).
The Laplacian of $\tilde{G}$ is a $(r+m) \times (r+m)$ matrix.
Now we consider the polynomial
\bq
 {\mathcal W}(x_1,...,x_n,z_1,...,z_m) & = & \det \; L\left(\tilde{G}\right)[r+1,...,r+m].
\eq
${\mathcal W}$ is a polynomial of degree $r=n-l+1$ in the variables $x_i$ and $z_j$.
We can expand ${\mathcal W}$ in polynomials homogeneous in the variables $z_j$:
\bq
 {\mathcal W} & = & {\mathcal W}^{(0)} + {\mathcal W}^{(1)} + {\mathcal W}^{(2)} + ... + {\mathcal W}^{(m)},
\eq
where ${\mathcal W}^{(k)}$ is homogeneous of degree $k$ in the variables $z_j$.
We further write
\bq
 {\mathcal W}^{(k)} & = &
 \sum\limits_{(j_1,...,j_k)} {\mathcal W}^{(k)}_{(j_1,...,j_k)}(x_1,...,x_n) \; z_{j_1} ... z_{j_k}.
\eq
The sum is over all indices with $1\le j_1<...<j_j \le m$.
The ${\mathcal W}^{(k)}_{(j_1,...,j_k)}$ are homogeneous polynomials of degree $r-k$ in the variables $x_i$.
For ${\mathcal W}^{(0)}$ and ${\mathcal W}^{(1)}$ one finds
\bq
\label{laplacian_W0_W1}
 {\mathcal W}^{(0)} = 0,
 & &
 {\mathcal W}^{(1)} = {\mathcal K}(x_1,...,x_n) \; \sum\limits_{j=1}^m z_j,
\eq
therefore
\bq
\label{laplacian_U}
 {\mathcal U} & = & x_1 ... x_n \; {\mathcal W}^{(1)}_{(j)}\left(\frac{1}{x_1},...,\frac{1}{x_n}\right),
\eq
for any $j\in\{1,...,m\}$.
${\mathcal F}_0$ is related to ${\mathcal W}^{(2)}$:
\bq
\label{laplacian_F_0}
 {\mathcal F}_0 & = & 
   x_1 ... x_n \sum\limits_{(j,k)} 
   \left( \frac{p_j \cdot p_k}{\mu^2} \right)
   \cdot 
   {\mathcal W}^{(2)}_{(j,k)}\left(\frac{1}{x_1},...,\frac{1}{x_n}\right).
\eq
The proof of eqs.~(\ref{laplacian_W0_W1})-(\ref{laplacian_F_0}) follows from the
all-minors matrix-tree theorem. 
The all-minors matrix-tree theorem states
\bq
{\mathcal W}(x_1,...,x_n,z_1,...,z_m) & = &
 \sum\limits_{F\in {\mathcal T}_m^{I,I}(\tilde{G})} \;\; \prod\limits_{e_j \in F} a_j,
\eq
with $I=(r+1,...,r+m)$ and $a_j=x_j$ if $e_j$ is an internal edge or $a_j=z_{j-n}$ if $e_j$ is an external edge.
The sum is over all $m$-forests of $\tilde{G}$, such that each tree in an $m$-forest contains
exactly one of the added vertices $v_{r+1}$, ..., $v_{r+m}$.
Each $m$-forest has $m$ connected components. The polynomial ${\mathcal W}^{(0)}$ by definition does not contain
any variable $z_j$. ${\mathcal W}^{(0)}$ would therefore correspond to forests where
all edges connecting the external vertices $e_{r+1}$, ..., $e_{r+m}$ have been cut.
The external vertices appear therefore as isolated vertices in the forest.
Such a forest must necessarily have more than $m$ connected components. This is a contradiction 
with the requirement of having exactly $m$ connected components and
therefore ${\mathcal W}^{(0)}=0$.
Next, we consider ${\mathcal W}^{(1)}$. Each term is linear in the variables $z_j$. 
Therefore $(m-1)$ vertices of the added vertices $v_{r+1}$, ..., $v_{r+m}$ appear as isolated vertices
in the $m$-forest. The remaining added vertex is connected to a spanning tree of $G$.
Summing over all possibilities one sees that ${\mathcal W}^{(1)}$ is given by the product of
$(z_1+...+z_m)$ with the Kirchhoff polynomial of $G$.
Finally we consider ${\mathcal W}^{(2)}$.
Here, $(m-2)$ of the added vertices appear as isolated vertices. The remaining two are connected to
a spanning $2$-forest of the graph $G$, one to each tree of the $2$-forest.
Summing over all possibilities one obtains eq.~(\ref{laplacian_F_0}).

Eq.~(\ref{laplacian_U}) and eq.~(\ref{laplacian_F_0}) together with eq.~(\ref{def_F_0})
allow the computation of the first and second Symanzik polynomial
from the Laplacian of the graph $\tilde{G}$.
This provides a third method for the computation of the graph polynomials ${\mathcal U}$ and ${\mathcal F}$.

As an example we consider the double-box graph of fig.~\ref{fig1}.
We attach an additional vertex to every external line.
\begin{figure}
\begin{center}
\includegraphics[bb= 110 605 310 730]{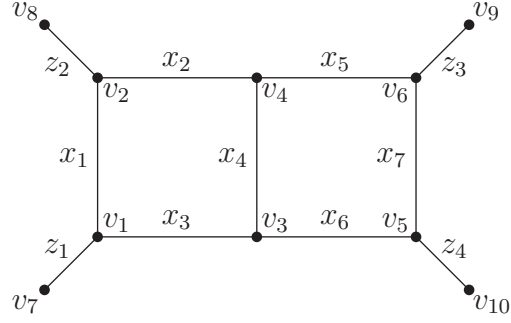}
\end{center}
\caption{\label{fig8} The labelling of the vertices and the Feynman parameters for the ``double box''-graph.}
\end{figure}    
Fig.~\ref{fig8} shows the labelling of the vertices and the Feynman parameters for the graph $\tilde{G}$.
The Laplacian of $\tilde{G}$ is a $10 \times 10$-matrix. We are interested in the minor, where 
-- with the labelling of fig.~\ref{fig8} -- we
delete the rows and columns $7$, $8$, $9$ and $10$.
The determinant of this minor reads
\bq
\lefteqn{
{\mathcal W} = \det \; L[7,8,9,10]
 = } & & 
 \nonumber \\
 & = &
 \left| \begin{array}{cccccc}
 x_1+x_3+z_1 & -x_1 & -x_3 & 0 & 0 & 0 \\
 -x_1 & x_1+x_2+z_2 & 0 & -x_2 & 0 & 0 \\
 -x_3 & 0 & x_3+x_4+x_6 & -x_4 & -x_6 & 0 \\
 0 & -x_2 & -x_4 & x_2+x_4+x_5 & 0 & -x_5 \\
 0 & 0 & -x_6 & 0 & x_6+x_7+z_4 & -x_7 \\
 0 & 0 & 0 & -x_5 & -x_7 & x_5+x_7+z_3 \\
 \end{array} \right|
 \nonumber \\
 & = &
 {\mathcal W}^{(1)} + {\mathcal W}^{(2)} + {\mathcal W}^{(3)}+ {\mathcal W}^{(4)}. 
\eq
For the polynomials ${\mathcal W}^{(1)}$ and ${\mathcal W}^{(2)}$ one finds 
\bq
\lefteqn{
{\mathcal W}^{(1)} = 
 \left( z_1 + z_2 + z_3 + z_4 \right) 
} & & 
 \nonumber \\
 & &
 \left(
          x_1 x_2 x_3 x_5 x_6 
        + x_1 x_2 x_3 x_5 x_7 
        + x_1 x_2 x_3 x_6 x_7 
        + x_1 x_2 x_4 x_5 x_6 
        + x_1 x_2 x_4 x_5 x_7 
 \right. \nonumber \\
 & & \left.
        + x_1 x_2 x_4 x_6 x_7 
        + x_1 x_2 x_5 x_6 x_7 
        + x_1 x_3 x_4 x_5 x_6 
        + x_1 x_3 x_4 x_5 x_7  
        + x_1 x_3 x_4 x_6 x_7 
 \right. \nonumber \\
 & & \left.
        + x_1 x_3 x_5 x_6 x_7 
        + x_2 x_3 x_4 x_5 x_6 
        + x_2 x_3 x_4 x_5 x_7  
        + x_2 x_3 x_4 x_6 x_7 
        + x_2 x_3 x_5 x_6 x_7  
 \right),
 \nonumber \\
\lefteqn{
{\mathcal W}^{(2)} =
          (z_1 + z_4) (z_2 + z_3) x_2 x_3 x_5 x_6
 } & & \nonumber \\
 & &  
     + (z_1 + z_2) (z_4 + z_3) 
      \left( 
           x_1 x_2 x_3 x_7 
         + x_1 x_2 x_4 x_7 
         + x_1 x_2 x_6 x_7 
         + x_1 x_3 x_4 x_7 
         + x_1 x_3 x_5 x_7 
         + x_1 x_4 x_5 x_6  
 \right. \nonumber \\
 & & \left.
         + x_1 x_4 x_5 x_7 
         + x_1 x_4 x_6 x_7 
         + x_1 x_5 x_6 x_7 
         + x_2 x_3 x_4 x_7 
      \right)
 \nonumber \\
 & &
         + z_1 (z_2 + z_3 + z_4) x_2
           \left( 
           x_3 x_5 x_7 
         + x_4 x_5 x_6 
         + x_4 x_5 x_7 
         + x_4 x_6 x_7 
         + x_5 x_6 x_7 
           \right)
 \nonumber \\
 & &
         + z_2 (z_1 + z_3 + z_4) x_3 
           \left(
           x_2 x_6 x_7  
         + x_4 x_5 x_6  
         + x_4 x_5 x_7 
         + x_4 x_6 x_7 
         + x_5 x_6 x_7 
           \right)
 \nonumber \\
 & &
         + z_3 (z_1 + z_2 + z_4) x_6
           \left(
           x_1 x_2 x_3  
         + x_1 x_2 x_4  
         + x_1 x_3 x_4  
         + x_1 x_3 x_5 
         + x_2 x_3 x_4 
           \right)
 \nonumber \\
 & &
         + z_4 (z_1 + z_2 + z_3) x_5 
           \left(
           x_1 x_2 x_3  
         + x_1 x_2 x_4 
         + x_1 x_2 x_6
         + x_1 x_3 x_4 
         + x_2 x_3 x_4  
          \right).
\eq
With the help of eq.~(\ref{laplacian_U}) and eq.~(\ref{laplacian_F_0})
and using the kinematical specifications of eq.~(\ref{specification_double_box}) we recover 
${\mathcal U}$ and ${\mathcal F}$ of eq.~(\ref{result_U_and_F_double_box}).

We would like to make a few remarks: The polynomial ${\mathcal W}$ is obtained from the determinant
of the matrix $L=L\left(\tilde{G}\right)[r+1,...,r+m]$. This matrix was constructed by attaching
one additional vertex to each external line. 
Then one deletes the rows and columns corresponding to the newly added vertices
from the Laplacian of the graph obtained in this way.
There are two alternative ways to arrive at the same matrix $L$:

The first alternative consists in attaching to the original graph $G$ a single new vertex $v_{r+1}$,
which connects to all external lines. This defines a new graph $\hat{G}$, which by construction no longer
has any external lines. 
As before we associate variables $z_1$, ..., $z_m$ to the edges connected to $v_{r+1}$.
\begin{figure}
\begin{center}
\includegraphics[bb= 135 635 465 720]{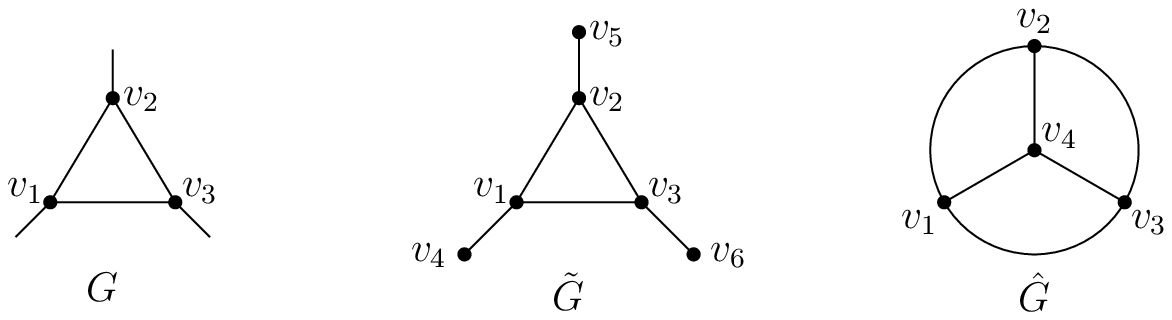}
\end{center}
\caption{\label{fig11} The left picture shows a one-loop graph with three external legs.
The middle pictures shows the associated graph $\tilde{G}$, where three additional vertices have been
attached to the external legs.
The right picture shows the graph $\hat{G}$ associated to $G$, where all external legs have been joined
in one additional vertex.}
\end{figure}    
Fig.~\ref{fig11} shows an example for the graphs $\tilde{G}$ and $\hat{G}$ associated to a
one-loop graph with three external legs.
The Laplacian of $\hat{G}$ is a $(r+1)\times(r+1)$-matrix.
It is easy to see that
\bq
 L & = & L\left(\hat{G}\right)[r+1].
\eq
From eq.~(\ref{matrix_tree_theorem}) we see that the polynomial ${\mathcal W}$ is nothing else than
the Kirchhoff polynomial of the graph $\hat{G}$:
\bq
\label{relation_W_K}
 {\mathcal W}(G) & = & {\mathcal K}\left( \hat{G} \right) 
 = \det L\left(\hat{G}\right)[j],
\eq
where $j$ is any number between $1$ and $r+1$.

For the second alternative one starts from the Laplacian of the original graph $G$. Let $\pi$ be a 
permutation of $(1,...,r)$. We consider the diagonal matrix
$\mbox{diag}\left(z_{\pi(1)},...,z_{\pi(r)}\right)$.
We can choose the permutation $\pi$ such that
\bq
 \pi(i) & = & j, \;\;\;\mbox{if the external line $j$ is attached to vertex $v_i$.}
\eq
We then have
\bq
 L & = & L\left( G \right) 
       + \left.\mbox{diag}\left(z_{\pi(1)},...,z_{\pi(r)}\right) \right|_{z_{m+1}=...=z_r=0}.
\eq


\section{Deletion and contraction properties}
\label{sect_deletion_and_contraction}

In this section we study two operations on a graph: the deletion of an edge and the contraction of an edge.
This leads to a recursive algorithm for the calculation of the graph polynomials ${\mathcal U}$ and ${\mathcal F}$.
In addition we discuss the multivariate Tutte polynomial and Dodgson's identity.

In graph theory an edge is called a bridge, if the deletion of the edge increases
the number of connected components. 
In the physics literature the term ``one-particle-reducible'' is used for a connected graph 
containing at least one bridge.
The contrary is called ``one-particle-irreducible'', i.e. a connected graph containing no bridges.
\begin{figure}
\begin{center}
\includegraphics[bb= 125 655 262 720]{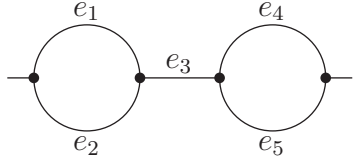}
\end{center}
\caption{\label{fig9} 
A one-particle-reducible graph: The edge $e_3$ is called a bridge. Deleting $e_3$ results in two
connected components.
}
\end{figure}    
Fig.~\ref{fig9} shows an example. The edge $e_3$ is a bridge, while the edges $e_1$, $e_2$, $e_4$ and
$e_5$ are not bridges.
Note that all edges of a tree graph are bridges.
An edge which is neither a bridge nor a self-loop is called a regular edge.
For a graph $G$ and a regular edge $e$ we define
\bq
G/e & & \mbox{to be the graph obtained from $G$ by contracting the regular edge $e$,} \nonumber \\
G- e & & \mbox{to be the graph obtained from $G$ by deleting the regular edge $e$.}
\eq
\begin{figure}
\begin{center}
\includegraphics[bb= 125 655 425 720]{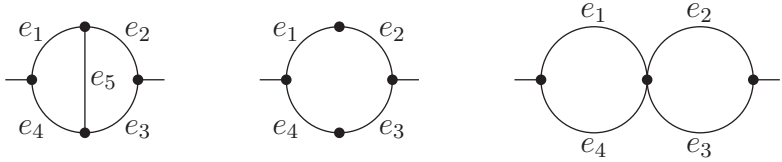}
\end{center}
\caption{\label{fig10} The left picture shows the graph $G$ of the two-loop two-point function.
The middle picture shows the graph $G-e_5$, where edge $e_5$ has been deleted.
The right picture shows the graph $G/e_5$, where the two vertices connected to $e_5$ have been joined and the edge
$e_5$ has been removed.
}
\end{figure}    
Fig.~\ref{fig10} shows an example. 
If the graph $G$ has loop number $l$ it follows that $G-e$ has loop number $(l-1)$, while
$G/e$ has loop number $l$.
This follows easily from the formula $l=n-r+1$ for a connected graph: $G-e$ has one edge less, but the
same number of vertices, while $G/e$ has one edge and one vertex less.

Let us now study the behaviour of the Laplacian under these operations. 
Under deletion the Laplacian behaves as
\bq
 L\left(G- e_k\right) & = & \left. L(G) \right|_{x_k=0},
\eq
i.e. the Laplacian of the graph $G-e_k$ is obtained from the Laplacian of the graph $G$ by setting the variable
$x_k$ to zero.
The behaviour of the Laplacian under contraction is slightly more complicated:
As before we consider a graph with $r$ vertices.
Assume that edge $e_k$ connects the vertices $v_a$ and $v_r$.
The Laplacian $L(G/ e_k)$ is then a $(r-1) \times (r-1)$-matrix with entries
\bq
 L\left(G/ e_k\right)_{ij}
 & = & \left\{ \begin{array}{ll}
   L(G)_{aa}+L(G)_{rr}+L(G)_{ar}+L(G)_{ra}, & \mbox{if $i=j=a$,} \\
   L(G)_{aj}+L(G)_{rj},      & \mbox{if $i=a$, $j\neq a$,} \\
   L(G)_{ia}+L(G)_{ir},      & \mbox{if $j=a$, $i\neq a$,} \\
   L(G)_{ij},                & \mbox{otherwise.} \\
  \end{array} \right.
\eq
Therefore the Laplacian of $L(G/ e_k)$ is identical to the minor $L(G)[r]$ except for the row and column $a$.
The choice that the edge $e_k$ is connected to the last vertex $v_r$ was merely made to keep the notation simple.
If the edge connects the vertices $v_a$ and $v_b$ with $a<b$ one deletes from $L(G)$ row and column $b$ and modifies
row and column $a$ analogously to the formula above with $b$ substituted for $r$.
In particular we have \cite{Godsil:2001}
\bq
 L\left(G/ e_k\right)[a] & = & L\left(G\right)[a,b].
\eq
The deletion/contraction operations can be used for a recursive definition of the graph polynomials.
For any regular edge $e_k$ we have
\bq
\label{recursion_U_F0}
 {\mathcal U}(G) & = & {\mathcal U}(G/e_k) + x_k {\mathcal U}(G-e_k), 
 \nonumber \\
 {\mathcal F}_0(G) & = & {\mathcal F}_0(G/e_k) + x_k {\mathcal F}_0(G-e_k).
\eq
The recursion terminates when all edges are either bridges or self-loops.
This is then a graph, which can be obtained from a tree graph by attaching self-loops to some
vertices. 
These graphs are called terminal forms.
If a terminal form has $r$ vertices and $l$ (self-) loops, 
then there are $(r-1)$ ``tree-like'' propagators, where the momenta flowing through these propagators
are linear combinations of the external momenta $p_i$ alone and independent of the independent loop momenta $k_j$.
The momenta of the remaining $l$ propagators are on the other hand independent of the external momenta 
and can be taken as the independent loop momenta $k_j$, $j=1,...,l$.
Let us agree that we label the $(r-1)$ ``tree-like'' edges from $1$ to $r-1$, and the remaining $l$
edges by $r$, ..., $n$ with $n=r+l-1$.
We further denote the momentum squared flowing through edge $j$ by $q_j^2$.
For a terminal form we have
\bq
\label{terminus_U_F0}
 {\mathcal U} = x_{r} ... x_{n},
 & &
 {\mathcal F}_0 = x_{r} ... x_{n} \sum\limits_{j=1}^{r-1} x_j \left( \frac{-q_j^2}{\mu^2} \right).
\eq
In the special case that the terminal form is a tree graph, this reduces to
\bq
 {\mathcal U} = 1,
 & &
 {\mathcal F}_0 = \sum\limits_{j=1}^{r-1} x_j \left( \frac{-q_j^2}{\mu^2} \right).
\eq
The Kirchhoff polynomial has for any regular edge the recursion relation
\bq
\label{recursion_K}
 {\mathcal K}(G) & = & x_k {\mathcal K}(G/e_k) + {\mathcal K}(G-e_k).
\eq
Note that the factor $x_k$ appears here in combination with the contracted graph $G/e_k$.
The recursion ends again on terminal forms. For these graphs we have with the 
conventions as above
\bq
\label{terminus_K}
 {\mathcal K} = x_1 ... x_{r-1}.
\eq
The recursion relations eq.~(\ref{recursion_U_F0}) and eq.~(\ref{recursion_K}) are proven with 
the help of the formulae, which express the polynomials ${\mathcal U}$ and ${\mathcal K}$ in terms of
spanning trees. For ${\mathcal F}_0$ one uses the corresponding formula, which expresses this polynomial
in terms of spanning $2$-forests.
As an example consider the polynomial ${\mathcal U}$ and the set of all spanning trees.
This set can be divided into two sub-sets: the first sub-set is given by the spanning trees, which contain the
edge $e_k$, while the second subset is given by those which do not.
The spanning trees in the first sub-set are in one-to-one correspondence with the spanning trees of $G/e_k$,
the relation is given by contracting and decontracting the edge $e_k$.
The second subset is identical to the set of spanning trees of $G-e_k$. The graphs $G$ and $G-e_k$ differ by the edge
$e_k$ which has been deleted, therefore the explicit factor $x_k$ in front of ${\mathcal U}(G-e_k)$.

Eq.~(\ref{recursion_U_F0}) and eq.~(\ref{terminus_U_F0}) together with eq.~(\ref{def_F_0})
provide a fourth method for the computation of the graph polynomials ${\mathcal U}$ and ${\mathcal F}$.

We now look at a generalisation of the Kirchhoff polynomial satisfying a recursion relation similar
to eq.~(\ref{recursion_K}).
For a graph $G$ -- not necessarily connected -- we denote by ${\mathcal S}$ the set of all spanning sub-graphs of $G$, i.e. sub-graphs $H$
of $G$, which contain all vertices of $G$. It is not required that a spanning sub-graph is a forest or a tree.
As before we associate to each edge $e_i$ a variable $x_i$. We will need one further formal variable $q$. 
We recall that the loop number of a graph $G$ with $n$ internal edges and $r$ vertices is given by
\bq
 l & = & n - r + k,
\eq
where $k$ is the number of connected components of $G$.
We can extend the definition of the deletion and contraction properties to edges which are not regular.
If the edge $e$ is a self loop, we define the contracted graph $G/e$ to be identical to $G-e$.
The multivariate Tutte polynomial is defined by \cite{Sokal:2005aa}
\bq
\label{def_Tutte}
{\mathcal Z}\left(q,x_1,...,x_n\right)
 & = &
 \sum\limits_{H \in {\mathcal S}} q^{k(H)}
 \prod\limits_{e_i \in H} x_i.
\eq
It is a polynomial in $q$ and $x_1$, ..., $x_n$.
The multivariate Tutte polynomial generalises the standard
Tutte polynomial \cite{Tutte:1947,Tutte:1954,Tutte:1967,Ellis-Monaghan:2008aa,Ellis-Monaghan:2008ab,Krajewski:2008aa},
which is a polynomial in two variables.
For the multivariate Tutte polynomial we have the recursion relation
\bq
\label{recursion_Z}
 {\mathcal Z}(G) & = & x_k {\mathcal Z}(G/e_k) + {\mathcal Z}(G-e_k),
\eq
where $e_k$ is any edge, not necessarily regular.
The terminal forms are graphs which consists solely of vertices without any edges.
For a graph with $r$ vertices and no edges one has
\bq
 {\mathcal Z} & = & q^r.
\eq
The multivariate Tutte polynomial starts as a polynomial in $q$ with $q^k$ if $G$ is a graph with $k$
connected components.
If the graph $G$ is not connected we write $G=(G_1,...,G_k)$, where $G_1$ to $G_k$ are the connected
components. For a disconnected graph the multivariate Tutte polynomial factorises:
\bq
 {\mathcal Z}\left(G\right)
 & = &
 {\mathcal Z}\left(G_1\right) ... {\mathcal Z}\left(G_k\right).
\eq
Some examples for the multivariate Tutte polynomial are
\bq
{\mathcal Z}\left(
\begin{picture}(20,10)(0,2)
 \Vertex(2,5){2}
 \Vertex(18,5){2}
 \Line(2,5)(18,5)
\end{picture}
\right) 
 & = & q x + q^2,
\nonumber \\
{\mathcal Z}\left(
\begin{picture}(20,10)(0,3)
 \Vertex(10,2){2}
 \CArc(10,8)(6,0,360)
\end{picture}
\right) 
 & = & q \left( x + 1 \right),
\nonumber \\
{\mathcal Z}\left(
\begin{picture}(20,10)(0,3)
 \Vertex(4,8){2}
 \Vertex(16,8){2}
 \CArc(10,8)(6,0,360)
\end{picture}
\right) 
 & = & q \left( x_1 x_2 + x_1 + x_2 \right) + q^2.
\eq
If $G$ is a connected graph we recover the Kirchhoff polynomial ${\mathcal K}(G)$ from the Tutte
polynomial ${\mathcal Z}(G)$ by first taking the coefficient of the linear term in $q$ and then retaining only
those terms with the lowest degree of homogeneity in the variables $x_i$.
Expressed in a formula we have
\bq
 {\mathcal K}\left(x_1,...,x_n\right)
 & = & 
 \lim\limits_{\lambda \rightarrow 0} \; \lim\limits_{q\rightarrow 0} \;
 \lambda^{1-r} q^{-1} {\mathcal Z}\left(q,\lambda x_1,...,\lambda x_n\right).
\eq
To prove this formula one first notices that the definition in eq.~(\ref{def_Tutte})
of the multivariate Tutte polynomial is equivalent to
\bq
{\mathcal Z}\left(q,x_1,...,x_n\right)
 & = &
 q^r
 \sum\limits_{H \in {\mathcal S}} q^{l(H)}
 \prod\limits_{e_i \in H} \frac{x_i}{q}.
\eq
One then obtains
\bq
 \lambda^{1-r} q^{-1} {\mathcal Z}\left(q,\lambda x_1,...,\lambda x_n\right)
 & = &
 \sum\limits_{H \in {\mathcal S}} q^{k(H)-1} \lambda^{l(H)-k(H)+1}
 \prod\limits_{e_i \in H} x_i.
\eq
The limits $q\rightarrow 0$ and $\lambda \rightarrow 0$ select $k(H)=1$ and $l(H)=0$, hence
the sum over the spanning sub-graphs reduces to a sum over spanning trees and one recovers
the Kirchhoff polynomial.

At the end of this section we want to discuss 
Dodgson's identity\footnote{Dodgson's even more famous literary work contains the
novel ``Alice in wonderland'' which he wrote using the pseudonym
Lewis Carroll.}.
Dodgson's identity states that for any $n \times n$ matrix $A$ and integers $i$, $j$ with $1\le i,j \le n$ and $i \neq j$ 
one has \cite{Dodgson:1866,Zeilberger:1997}
\bq
\label{dodgson_identity}
 \det\left( A \right) \det\left( A[i,j] \right)
 & = & 
 \det\left( A[i] \right) \det\left( A[j] \right)
 -
 \det\left( A[i;j] \right) \det\left( A[j;i] \right).
\eq
We remind the reader that $A[i,j]$ denotes a $(n-2)\times(n-2)$ matrix obtained from $A$ by deleting
the rows and columns $i$ and $j$.
On the other hand $A[i;j]$ denotes a $(n-1)\times(n-1)$ matrix which is obtained from $A$ by deleting
the $i$-th row and the $j$-th column.
The identity in eq.~(\ref{dodgson_identity}) 
has an interesting application towards graph polynomials: 
Let $e_a$ and $e_b$ be two regular edges of a graph $G$, which share one common vertex.
Assume that the edge $e_a$ connects the vertices $v_i$ and $v_k$, while the edge $e_b$ connects the vertices
$v_j$ and $v_k$. The condition $i \neq j$ ensures that after contraction of one edge the other edge
is still regular.
(If we would allow $i=j$ we have a multiple edge 
and the contraction of one edge leads to a self-loop for the other edge.)
For the Kirchhoff polynomial of the graph $G-e_a-e_b$ we have
\bq
 {\mathcal K}\left( G-e_a-e_b \right) & = & \det L\left( G-e_a-e_b\right)[k].
\eq
Let us now consider the Kirchhoff polynomials of the graphs $G/e_a-e_b$ and $G/e_b-e_a$. One finds
\bq
 {\mathcal K}\left( G/e_a-e_b \right) & = & \det L\left( G-e_a-e_b\right)[i,k],
 \nonumber \\
 {\mathcal K}\left( G/e_b-e_a \right) & = & \det L\left( G-e_a-e_b\right)[j,k].
\eq
Here we made use of the fact that the operations of contraction and deletion commute
(i.e. $G/e_a-e_b=(G-e_b)/e_a$) as well as of the fact that the variable $x_a$ occurs in the Laplacian
of $G$ only in rows and columns $i$ and $k$, therefore 
$L\left( G-e_b\right)[i,k] = L\left( G-e_a-e_b\right)[i,k]$.
Finally we consider the Kirchhoff polynomial of the graph $G/e_a/e_b$, for which one finds
\bq
 {\mathcal K}\left( G/e_a/e_b \right) & = & \det L\left( G-e_a-e_b\right)[i,j,k].
\eq
The Laplacian of any graph is a symmetric matrix, therefore
\bq
 \det L( G-e_a-e_b )[i,k;j,k] & = & \det L( G-e_a-e_b )[j,k;i,k].
\eq
We can now apply Dodgson's identity to the matrix $L( G-e_a-e_b )[k]$.
Using the fact that $L\left( G-e_a-e_b\right)[i,k;j,k]=L\left( G \right)[i,k;j,k]$ one finds \cite{Stembridge:1998}
\bq
\label{Dodgson_K}
 {\mathcal K}\left( G/e_a- e_b \right) {\mathcal K}\left( G/e_b- e_a \right)
 -
 {\mathcal K}\left( G- e_a-e_b \right) {\mathcal K}\left( G/e_a/e_b \right)
 = 
 \left( \det L\left( G \right)[i,k;j,k] \right)^2.
\eq
This equation shows that the expression on the left-hand side factorises into a square.
The expression on the right-hand side can be re-expressed using the all-minors matrix-tree theorem
as a sum over $2$-forests, such that the vertex $v_k$ is contained in one tree of the 
$2$-forest, while the vertices $v_i$ and $v_j$ are both contained in the other tree.

Expressed in terms of the first Symanzik polynomial we have
\bq
\label{factorisation_U_U}
 {\mathcal U}\left( G/e_a- e_b \right) {\mathcal U}\left( G/e_b- e_a \right)
 -
 {\mathcal U}\left( G- e_a-e_b \right) {\mathcal U}\left( G/e_a/e_b \right)
 & = &
 \left( \frac{\Delta_1}{x_a x_b} \right)^2.
\eq
The expression $\Delta_1$ is given by
\bq
\Delta_1 & = &
\sum\limits_{F\in {\mathcal T}_2^{(i,k),(j,k)}} \; \prod\limits_{e_t \notin F} x_t.
\eq
The sum is over all $2$-forests $F=(T_1,T_2)$ of $G$ such that $v_i, v_j \in T_1$ and $v_k \in T_2$.
Note that each term of $\Delta_1$ contains $x_a$ and $x_b$.
The factorisation property of eq.~(\ref{factorisation_U_U}) plays a crucial role
in the algorithms of refs.~\cite{Brown:2008,Brown:2009a,Brown:2009b}.

A factorisation formula similar to eq.~(\ref{factorisation_U_U}) can be derived for an expression containing
both the first Symanzik polynomial ${\mathcal U}$ and the polynomial ${\mathcal F}_0$.
As before we assume that $e_a$ and $e_b$ are two regular edges of a graph $G$, which share one common vertex.
The derivation uses the results of sect.~\ref{sect_matrix_tree_theorem}
and starts from eq.~(\ref{Dodgson_K}) for the graph $\hat{G}$ associated to $G$. 
Eq.~(\ref{relation_W_K}) relates then the Kirchhoff 
polynomial of $\hat{G}$ to the ${\mathcal W}$-polynomial of $G$.
The ${\mathcal W}$-polynomial is then expanded in powers of $z$. The lowest order terms reproduce eq.~(\ref{factorisation_U_U}).
The next order yields
\bq
 {\mathcal U}\left( G/e_a- e_b \right) {\mathcal F}_0\left( G/e_b- e_a \right)
 -
 {\mathcal U}\left( G- e_a-e_b \right) {\mathcal F}_0\left( G/e_a/e_b \right)
&&
\\
 +
 {\mathcal F}_0\left( G/e_a- e_b \right) {\mathcal U}\left( G/e_b- e_a \right)
 -
 {\mathcal F}_0\left( G- e_a-e_b \right) {\mathcal U}\left( G/e_a/e_b \right)
 & = & 
 2 \left( \frac{\Delta_1}{x_a x_b} \right) \left( \frac{\Delta_2}{x_a x_b} \right).
 \nonumber 
\eq
The quantity $\Delta_2$ appearing on the right-hand side is obtained from the
all-minors matrix-tree theorem. We can express this quantity in terms of spanning three-forests of $G$ as follows:
Let us denote by ${\mathcal T}_3^{((i,j),\bullet,k)}$ the set of spanning 
three-forests $(T_1,T_2,T_3)$ of $G$ such that
$v_i, v_j \in T_1$ and $v_k \in T_3$.
Similar we denote by ${\mathcal T}_3^{(i,j,k)}$ the set of spanning 
three-forests $(T_1,T_2,T_3)$ of $G$ such that
$v_i \in T_1$, $v_j \in T_2$ and $v_k \in T_3$.
Then
\bq
\Delta_2 & = &
\sum\limits_{(T_1,T_2,T_3)\in {\mathcal T}_3^{(i,j,k)}} \; 
 \sum\limits_{v_c \in T_1, v_d \in T_2} \; \left( \frac{p_c \cdot p_d}{\mu^2} \right)
 \prod\limits_{e_t \notin (T_1,T_2,T_3)} \; x_t
 \nonumber \\
 & & 
-\sum\limits_{(T_1,T_2,T_3)\in {\mathcal T}_3^{((i,j),\bullet,k)}} \; 
 \sum\limits_{v_c, v_d \in T_2} \; \left( \frac{p_c \cdot p_d}{\mu^2} \right)
 \prod\limits_{e_t \notin (T_1,T_2,T_3)} \; x_t.
\eq
In this formula we used the convention that the momentum $p_j$ equals zero if no external leg is 
attached to vertex $v_j$.
Expanding the ${\mathcal W}$-polynomial to order $z^4$ we have terms of order $z^2$ squared as well as terms
which are products of order $z$ with order $z^3$. We are interested in an expression which arises 
from terms of order $z^2$ squared alone. In this case we obtain a factorisation formula only for special
kinematical configurations.
If for all external momenta one has
\bq
\label{condition_external_momenta}
 \left( p_{i_1} \cdot p_{i_2} \right) \cdot \left( p_{i_3} \cdot p_{i_4} \right)
 & = & 
 \left( p_{i_1} \cdot p_{i_3} \right) \cdot \left( p_{i_2} \cdot p_{i_4} \right),
 \;\;\;\;\;\;
 i_1, i_2, i_3, i_4 \in \{1,...,m\}
\eq
then
\bq
 {\mathcal F}_0\left( G/e_a- e_b \right) {\mathcal F}_0\left( G/e_b- e_a \right)
 -
 {\mathcal F}_0\left( G- e_a-e_b \right) {\mathcal F}_0\left( G/e_a/e_b \right)
 & = &
 \left( \frac{\Delta_2}{x_a x_b} \right)^2.
\eq
Eq.~(\ref{condition_external_momenta}) is satisfied for example if all external momenta are collinear.
A second example is given by a three-point function. In the kinematical configuration where
\bq
 \left( p_1^2 \right)^2 + \left( p_2^2 \right)^2 + \left( p_3^2 \right)^2
 - 2 p_1^2 p_2^2 - 2 p_1^2 p_3^2 - 2 p_2^2 p_3^2 & = & 0,
\eq
eq.~(\ref{condition_external_momenta}) is satisfied. 


\section{Duality}
\label{sect_duality}

We have seen that the Kirchhoff polynomial ${\mathcal K}$ and the first Symanzik polynomial ${\mathcal U}$ 
of a graph $G$ with $n$ edges are related by the equations~(\ref{convert_U_K}):
\bq
 {\mathcal U}(x_1,...,x_n) = x_1 ... x_n \; {\mathcal K}\left(\frac{1}{x_1},...,\frac{1}{x_n}\right),
 & &
 {\mathcal K}(x_1,...,x_n) = x_1 ... x_n \; {\mathcal U}\left(\frac{1}{x_1},...,\frac{1}{x_n}\right).
 \nonumber 
\eq
In this section we will ask if one can find a graph $G^\ast$ with $n$ edges such that 
${\mathcal K}(G^\ast)= {\mathcal U}(G)$ and 
${\mathcal K}(G)= {\mathcal U}(G^\ast)$.
Such a graph $G^\ast$ will be called a dual graph of $G$.
In this section we will show that for a planar graph one can always construct a dual graph.
The dual graph of $G$ need not be unique, there might be several topologically distinct graphs $G^\ast$ fulfilling the above mentioned
relation.
In other words two topologically distinct graphs $G_1$ and $G_2$ both of them with $n$ edges can have the same Kirchhoff polynomial.

In this section we consider only graphs with no external lines.
If a graph contains an external line, we can convert it to a graph with no external lines by
attaching an additional vertex to the end of the external line.
This is the same operation as the one considered already in fig.~\ref{fig8}.
\begin{figure}
\begin{center}
\includegraphics[bb= 105 537 505 646,width=0.7\textwidth]{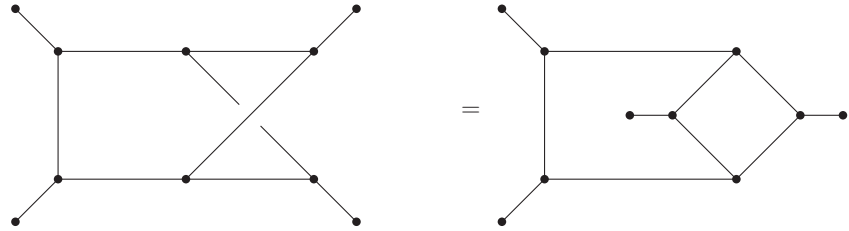}
\end{center}
\caption{\label{fig12} The ``crossed double-box''-graph can be drawn as a planar graph.}
\end{figure}    
A graph is called planar if it can be embedded in a plane without
crossings of edges.
We would like to note that the ``crossed double-box''-graph shown in fig.~\ref{fig12} is a planar graph. 
The right picture of fig.~\ref{fig12} shows how this graph can be drawn in the plane without any crossing of edges.

\begin{figure}
\begin{center}
\includegraphics[bb= 296 386 456 546, scale=0.5]{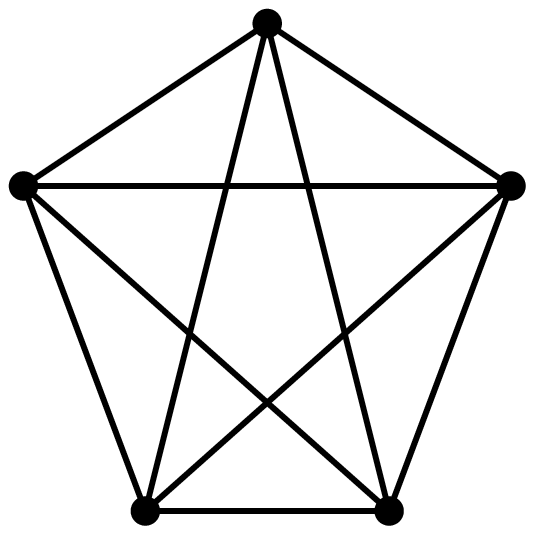}
\hspace*{15mm}
\includegraphics[bb= 296 386 433 546, scale=0.5]{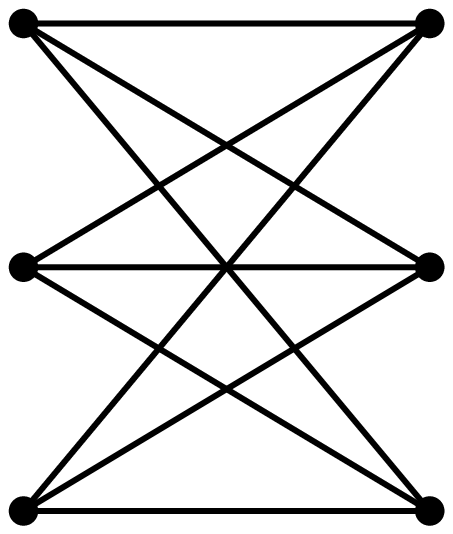}

~~~~$K_{5}$~~~~~~~~~~~~~~~~~~~~~~~~~~~~~~~~~~~$K_{3,\,3}$
\end{center}
\caption{\label{fig13} The 'smallest' non-planar graphs.
}
\end{figure}

Fig.~\ref{fig13} shows two examples of non-planar graphs. 
The first graph is the complete graph with five vertices $K_5$. The second example is denoted $K_{3,\,3}$.
A theorem states that a graph $G$ is planar
if and only if none of the graphs obtained from $G$ by a (possibly
empty) sequence of contractions of edges contains $K_{5}$ or $K_{3,\,3}$
as a sub-graph \cite{Kuratowski:1930,Wagner:1937,Diestel:2005}. 

\begin{figure}
\begin{center}
\includegraphics[bb= 296 386 502 499, scale=0.5]{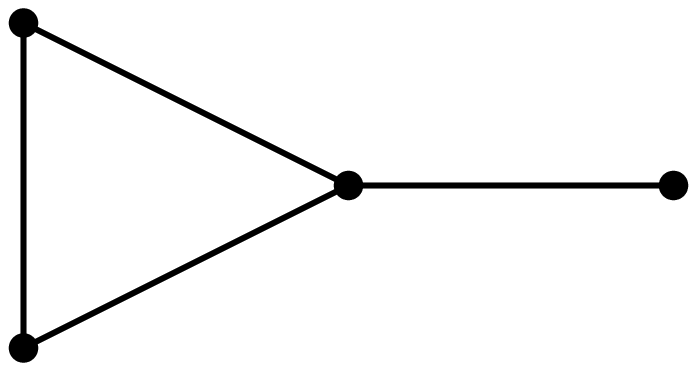}
\hspace*{15mm}
\includegraphics[bb= 296 345 450 447, scale=0.5]{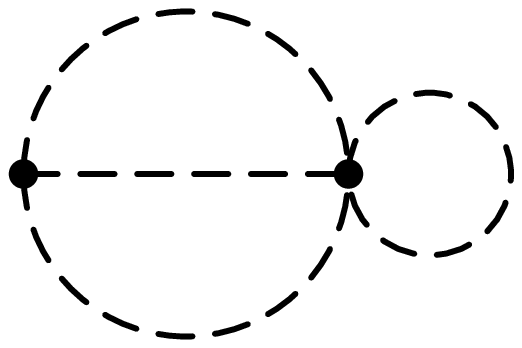}
\hspace*{15mm}
\includegraphics[bb= 69 333 316 482, scale=0.5]{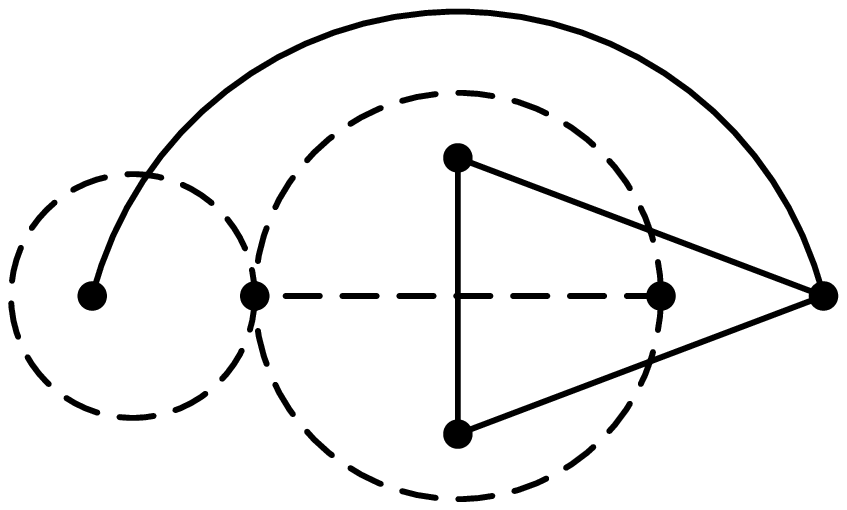}
\end{center}
\caption{\label{fig14} The first two pictures show a graph $G$ and its dual graph $G^\ast$.
The right picture shows the construction of $G^\ast$ from $G$ (or vice versa).
}
\end{figure}

Each planar graph $G$ has a dual graph $G^{\star}$ which
can be obtained as follows:
\begin{itemize}
\item Draw the graph $G$ in a plane, such that no edges intersect. In this
way, the graph divides the plane into open subsets, called faces.
\item Draw a vertex inside each face. These are the vertices of $G^{\star}$.
\item For each edge $e_{i}$ of $G$ draw a new edge $e_i^\ast$ connecting the two
vertices of the faces, which are separated by $e_{i}$. The new edges $e_i^\ast$
are the edges of $G^{\star}$.
\end{itemize}
An example for this construction is shown in fig.~\ref{fig14}.
We note from the construction of the dual graph, that for each external
edge in $G$ there is a self-loop in $G^{\star}$ and that for
each self-loop in $G$ there is an external edge in $G^{\star}$.

\begin{figure}
\begin{center}
\includegraphics[bb= 122 530 403 708]{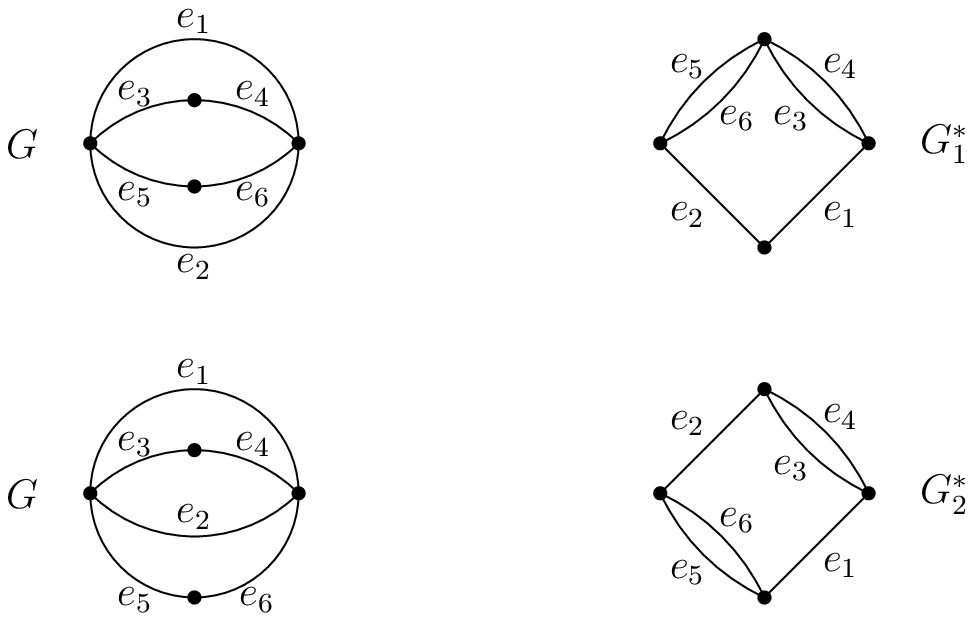}
\end{center}
\caption{\label{fig15} An example showing that different embeddings of a planar graph $G$ into the plane
yield different dual graphs $G^\ast_1$ and $G^\ast_2$.}
\end{figure}

If we now associate the variable $x_i$ to the edge $e_i$ of $G$ as well as to the edge $e_i^\ast$
of $G^\ast$ we have
\bq
{\mathcal K}(G^\ast)= {\mathcal U}(G),
&& 
{\mathcal K}(G)= {\mathcal U}(G^\ast).
\eq
It is important to note that the above construction of the dual graph
$G^{\star}$ depends on the way, how $G$ is drawn in the plane.
A given graph $G$ can have several topologically distinct dual graphs. 
These dual graphs have the same Kirchhoff polynomial. 
An example is shown in fig.~\ref{fig15}.
For this example one finds
\bq
& &
 {\mathcal K}(G) = {\mathcal U}(G^\ast_1) = {\mathcal U}(G^\ast_2)
 = 
 (x_1+x_2)(x_3+x_4)(x_5+x_6) + x_3 x_4 (x_5+x_6) + (x_3+x_4) x_5 x_6,
 \nonumber \\
& &
 {\mathcal U}(G) = {\mathcal K}(G^\ast_1) = {\mathcal K}(G^\ast_2)
 = 
 x_1 x_2 ( x_3 + x_4 + x_5 + x_6 ) + ( x_1 + x_2 ) ( x_3 + x_4 ) ( x_5 + x_6 ).
\eq


\section{Matroids}
\label{sect_matroids}

In this section we introduce the basic terminology of matroid theory.
We are in particular interested in cycle matroids.
A cycle matroid can be constructed from
a graph and it contains the information which is needed for
the construction of the Kirchhoff polynomial 
and therefore as well for the construction of the first Symanzik polynomial $\mathcal{U}$.
In terms of matroids we want to discuss the fact, that two different
graphs can have the same Kirchhoff polynomial. 
We have already encountered an example in fig.~\ref{fig15}.
We review a theorem on matroids which determines the classes of graphs whose Kirchhoff
polynomials are the same. For a detailed introduction to matroid theory
we refer to \cite{Oxley,Oxley:2003aa}.

\begin{figure}
\begin{center}
\includegraphics[bb= 285 588 420 697]{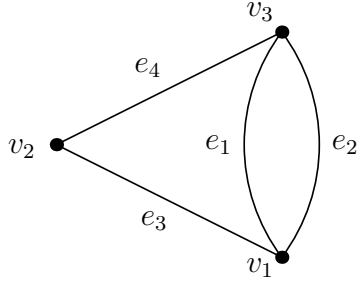}
\end{center}
\caption{\label{fig17} A graph $G$.}
\end{figure}

We introduce cycle matroids by an
example and consider the graph $G$ of fig.~\ref{fig17}.
The graph $G$ has three vertices $V=\left\{ v_{1},\, v_{2},\, v_{3}\right\}$
and four edges $E=\left\{ e_{1},\, e_{2},\, e_{3},\, e_{4}\right\}$.
The graph has five spanning trees given by the sets of
edges $\left\{ e_{1},\, e_{3}\right\}$, $\left\{ e_{1},\, e_{4}\right\}$,
$\left\{ e_{2},\, e_{3}\right\}$, $\left\{ e_{2},\, e_{4}\right\}$,
$\left\{ e_{3},\, e_{4}\right\}$, respectively. We obtain the Kirchhoff
polynomial ${\mathcal K}=x_{1}x_{3}+x_{1}x_{4}+x_{2}x_{3}+x_{2}x_{4}+x_{3}x_{4}$. 
The incidence matrix of a graph $G$ with $r$ vertices and $n$ edges is 
a $r \times n$-matrix $B=(b_{ij})$, defined by 
\bq
b_{ij} & = & \left\{ \begin{array}{l}
1,\textrm{\ensuremath{\;} if }e_{j}\textrm{ is incident to }v_{i},\\
0,\textrm{\ensuremath{\;} else.}\end{array}\right.
\eq
For the graph $G$ of fig.~\ref{fig17} the incidence matrix reads
\[
\begin{array}{cccc}
e_{1} & e_{2} & e_{3} & e_{4}
\end{array}
\]
\[
\left(\begin{array}{cccc}
1\, & \,1\, & \,1\, & 0\\
0 & 0 & 1 & 1\\
1 & 1 & 0 & 1
\end{array}\right),
\]
where we indicated that each column vector corresponds to one edge
of the graph. Let us focus on the set of these four column vectors.
We want to consider all subsets of these vectors which are linearly
independent over $\mathbb{Z}_2$. Obviously the set of all four vectors
and the set of any three of the given vectors are linearly dependent
over $\mathbb{Z}_2.$ Furthermore the first two columns corresponding
to $\left\{ e_{1},\, e_{2}\right\} $ are equal and therefore linearly
dependent. Hence the linearly independent subsets are all sets with
only one vector and all sets with two vectors, except for the just
mentioned one consisting of the first and second column. 
For each set of linearly independent vectors let us now write the set of the
corresponding edges. The set of all these sets shall be denoted ${\mathcalI}$.
We obtain
\bq
\label{example_independent_sets}
{\mathcalI} 
 & = & 
 \left\{ \emptyset, \, \left\{ e_{1}\right\}, \, \left\{ e_{2}\right\}, \,
         \left\{ e_{3}\right\}, \, \left\{ e_{4}\right\},\right. \nonumber \\
 &  & \left.\left\{ e_{1}, \, e_{3}\right\}, \, \left\{ e_{1}, \, e_{4}\right\}, \,
      \left\{ e_{2}, \, e_{3}\right\}, \, \left\{ e_{2}, \, e_{4}\right\}, \,
      \left\{ e_{3}, \, e_{4}\right\} \right\}.
\eq
The empty set is said to be independent and is included here
by definition. Let us make the important observation, that the sets
in ${\mathcalI}$ which have two elements, i.e. the maximal number
of elements, are exactly the sets of edges of the spanning trees of
the graph given above. 

The pair $(E, \, \mathcalI)$ consisting of the set of edges $E$ 
and the set of linearly independent sets $\mathcalI$ is an example of a matroid.
Matroids are defined as ordered pairs $(E,\,\mathcalI)$
where $E$ is a finite set, the ground set, and where $\mathcalI$
is a collection of subsets of $E$, called the independent sets,
fulfilling the following conditions:
\begin{enumerate}
\item $\emptyset\in\mathcalI.$
\item If $I\in\mathcalI$ and $I'\subseteq I,$ then $I'\in\mathcalI.$
\item If $I_{1}$ and $I_{2}$ are in $\mathcalI$ and $\left|I_{1}\right|<\left|I_{2}\right|,$
then there is an element $e$ of $I_{2}-I_{1}$ such that $I_{1}\cup e\in\mathcalI.$
\end{enumerate}
All subsets of $E$ which do not belong to $\mathcalI$ are called
dependent. 
The definition goes back to Whitney who wanted 
to describe the properties of linearly independent sets in an abstract way. 
In a similar way as a topology on a space
is given by the distinction between open and closed sets, a matroid
is given by deciding, which of the subsets of a ground set $E$ shall
be called independent and which dependent. 
A matroid can be defined on any kind of ground set, but if we choose $E$ to
be a set of vectors, we can see that the conditions for the independent
sets match with the well-known linear independence of vectors. 

Let us go through the three conditions. The first condition simply says
that the empty set shall be called independent. The second condition states that a
subset of an independent set is again an independent set. 
This is fulfilled for sets of linearly independent vectors as we already have seen in
the above example. 
The third condition is called the independence
augmentation axiom and may be clarified by an example. Let the sets
\[
I_{1}=\left\{ \left(\begin{array}{c}
1\\
0\\
0\\
0\end{array}\right),\,\left(\begin{array}{c}
0\\
1\\
0\\
0\end{array}\right)\right\} ,\quad I_{2}=\left\{ \left(\begin{array}{c}
1\\
0\\
0\\
0\end{array}\right),\,\left(\begin{array}{c}
0\\
1\\
1\\
1\end{array}\right),\,\left(\begin{array}{c}
1\\
0\\
0\\
1\end{array}\right)\right\} \]
be linearly independent sets of vectors, where $I_{2}$ has more elements
than $I_{1}$. $I_{2}-I_{1}$ is the set of vectors in $I_{2}$ which
do not belong to $I_{1}$. The set $I_{2}-I_{1}$ contains for example
$e=(1,\,0,\,0,\,1)^{T}$ and if we include this vector in $I_{1}$
then we obtain again a linearly independent set. The third condition states 
that such an $e$ can be found for any two independent sets with different
numbers of elements. 

The most important origins of examples of matroids are linear algebra
and graph theory. The cycle matroid (or polygon matroid)
of a graph $G$ is the matroid whose ground set $E$ is given by the
edges of $G$ and whose independent sets $\mathcalI$ are given
by the linearly independent subsets over ${\mathbb Z}_2$ of column vectors in the incidence
matrix of $G$. We can convince ourselves, that $\mathcalI$
fulfills the conditions layed out above.

Let us consider the bases or maximal independent sets
of a matroid $(E,\,\mathcalI).$ These are the sets in $\mathcalI$
which are not proper subsets of any other sets in $\mathcalI$.
The set of these bases of a matroid shall be denoted $\mathcal{B}$
and it can be defined by the following conditions:
\begin{enumerate}
\item $\mathcal{B}$ is non-empty.
\item If $B_{1}$ and $B_{2}$ are members of $\mathcal{B}$ and $x\in B_{1}-B_{2},$
then there is an element $y$ of $B_{2}-B_{1}$ such that $(B_{1}-x)\cup y\in\mathcal{B}.$
\end{enumerate}
One can show, that all sets in $\mathcal{B}$ have the same number
of elements. Furthermore $\mathcalI$ is uniquely determined by
$\mathcal{B}$: it contains the empty set and all subsets of members
of $\mathcal{B}$. 

Let $M=(E,\,\mathcalI)$ be the cycle matroid of a connected
graph $G$ and let $\mathcal{B}(M)$ be the set of bases. Then
one can show that $\mathcal{B}(M)$ consists of the sets of edges
of the spanning trees in $G$. In other words, $T$ is a spanning
tree of $G$ if and only if its set of edges are a basis in $\mathcal{B}(M)$.
We can therefore relate the Kirchhoff polynomial to the bases 
of the cycle matroid:
\bq
\label{eq:basis_generating}
\mathcal{K} & = & \sum_{B_{j}\in\mathcal{B}(M)}\; \prod_{e_{i}\in B_{j}}x_{i}.
\eq
The Kirchhoff polynomial of $G$ is called a 
basis generating polynomial of the matroid $M$ associated to $G$. 
The Kirchhoff polynomial allows us to read off the set of bases $\mathcal{B}(M)$.
Therefore two graphs without any self-loops have the same Kirchhoff polynomial if and
only if they have the same cycle matroid associated to them. 

Let us cure a certain ambiguity which is still left when we consider
cycle matroids and Kirchhoff polynomials and which comes from the
freedom in giving names to the edges of the graph. In the graph of
fig.~\ref{fig17} we could obviously choose different
names for the edges and their Feynman parameters, for example the
edge $e_{2}$ could instead be named $e_{3}$ and vice versa. As a
consequence we would obtain a different cycle matroid where compared
to the above one, $e_{3}$ takes the place of $e_{2}$ and vice versa.
Similarly, we would obtain a different Kirchhoff polynomial, where
the variables $x_{2}$ and $x_{3}$ are exchanged. Of course we are
not interested in any such different cases which simply result from
a change of variable names. Therefore it makes sense to consider classes
of isomorphic matroids and Kirchhoff polynomials. 

Let $M_{1}$ and $M_{2}$ be two matroids and let $E(M_{1})$ and
$E(M_{2})$ be their ground sets, respectively. The matroids $M_{1}$ and $M_{2}$
are called isomorphic if there is a bijection $\psi$ from
$E(M_{1})$ to $E(M_{2})$ such that $\psi(I)$ is an independent
set in $M_{2}$ if and only if $I$ is an independent set in $M_{1}$.
The mentioned interchange of $e_{2}$ and $e_{3}$ in the above example
would be such a bijection: $\psi(e_{2})=e_{3}$, $\psi(e_{3})=e_{2}$,
$\psi(e_{i})=e_{i}$ for $i=1,\,4.$ The new matroid obtained this
way is isomorphic to the above one and its independent sets are given by
interchanging $e_{2}$ and $e_{3}$ in $\mathcalI$ of eq.~(\ref{example_independent_sets}).
In the same sense we want to say that two Kirchhoff polynomials are
isomorphic if they are equal up to bijections on their sets of variables. 

\begin{figure}
\begin{center}
\includegraphics[bb= 185 582 518 702]{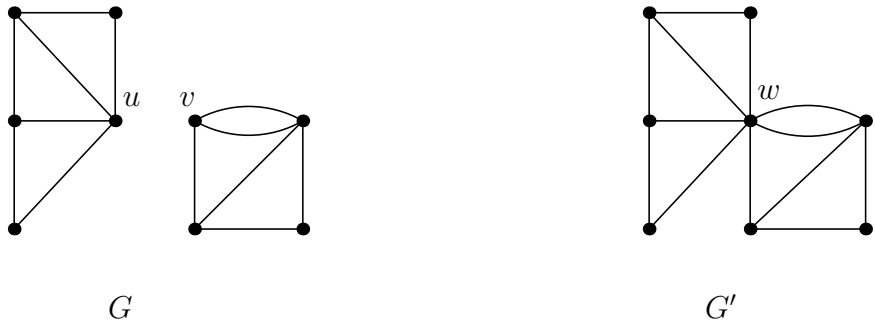}
\end{center}
\caption{Vertex identification and cleaving\label{fig:cleaving identification}}
\end{figure}

\begin{figure}
\begin{center}
\includegraphics[bb= 150 563 584 711]{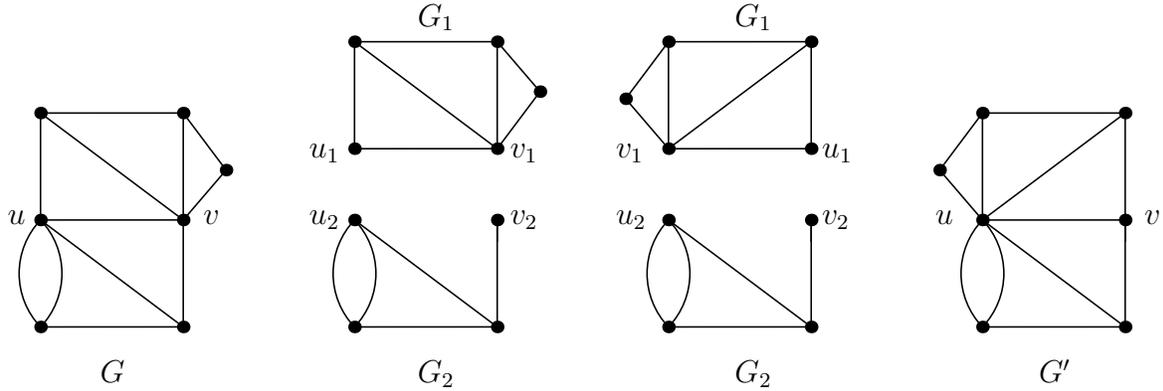}
\end{center}
\caption{Twisting about $u$ and $v$\label{fig:twisting}}
\end{figure}

Now let us come to the question when the Kirchhoff polynomials of
two different graphs are isomorphic, which means that after an appropriate
change of variable names they are equal. From the above discussion
and eq.~(\ref{eq:basis_generating}) it is now clear, that these
are exactly the cases where the cycle matroids of the graphs are isomorphic.
The question when two graphs have isomorphic cycle matroids was answered
in the following theorem of Whitney \cite{Whitney:1933aa} (also see \cite{Oxley:1986aa})
which was one of the foundational results of matroid theory: 

Let $G$ and $H$ be graphs having no isolated vertices. Then the
cycle matroids $M(G)$ and $M(H)$ are isomorphic if and only if
$G$ is obtained from $H$ after a sequence of the following three
transformations:
\begin{enumerate}
\item Vertex identification: Let $u$ and $v$ be vertices of distinct
components of a graph $G.$ Then a new graph $G'$ is obtained from
the identification of $u$ and $v$ as a new vertex $w$ in $G'$
(see the transition from $G$ to $G'$ in fig.~\ref{fig:cleaving identification}).
\item Vertex cleaving: Vertex cleaving is the reverse operation of
vertex identification, such that from cleaving at vertex $w$ in $G'$
we obtain $u$ and $v$ in distinct components of $G$ (see the transition
from $G'$ to $G$ in fig.~\ref{fig:cleaving identification}).
\item Twisting: Let $G$ be a graph which is obtained from two disjoint
graphs $G_{1}$ and $G_{2}$ by identifying the vertices $u_{1}$ of $G_{1}$
and $u_{2}$ of $G_{2}$ as a vertex $u$ of $G$ and by identifying
the vertices $v_{1}$ of $G_{1}$ and $v_{2}$ of $G_{2}$ as a vertex $v$
of $G$. Then the graph $G'$ is called the twisting of $G$ about
$\{u,\, v\}$ if it is obtained from $G_{1}$ and $G_{2}$ by identifying
instead $u_{1}$ with $v_{2}$ and $v_{1}$ with $u_{2}$ (see fig.~\ref{fig:twisting}).
\end{enumerate}
Proofs can be obtained from \cite{Oxley:1986aa,Whitney:1933aa,Truemper:1980aa,Oxley}. 
As a consequence of Whitney's theorem, the Kirchhoff polynomials $\mathcal{K}(G)$
and $\mathcal{K}(H)$ of the connected graphs $G$ and $H$, both without any self-loops, are isomorphic
if and only if $G$ is obtained from $H$ by a sequence of the above
three transformations. For the transformations of vertex identification
and vertex cleaving this is obvious from a well-known observation:
If two distinct components $G_{1}$ and $G_{2}$ are obtained from
$G$ after vertex cleaving, then $\mathcal{K}(G)$ is the product
of $\mathcal{K}(G_1)$ and $\mathcal{K}(G_2)$. Therefore any
other graph $G'$ obtained from $G_{1}$ and $G_{2}$ after vertex
identification has the same Kirchhoff polynomial 
$\mathcal{K}(G')=\mathcal{K}(G_1) \cdot \mathcal{K}(G_2)=\mathcal{K}(G)$.
The non-trivial part of the statement on the Kirchhoff polynomials
of $G$ and $H$ is the relevance of the operation of twisting. In
the initial example of fig.~\ref{fig15} the two graphs $G_1^\ast$ and $G_2^\ast$ can be obtained
from each other by twisting.

There is an alternative definition of a matroid: Instead of specifying the set ${\mathcalI}$
of independent sets a matroid can be defined by a rank function, which associates a non-negative
integer to every sub-set of the ground set.
The rank function has to satisfy for all $S,S' \subseteq E$ the following three conditions:
\begin{enumerate}
\item $\mbox{rk}(S) \le \left|S\right|$.
\item $S' \subset S$ implies $\mbox{rk}(S') \le \mbox{rk}(S)$.
\item $\mbox{rk}(S\cup S') + \mbox{rk}(S\cap S') \le \mbox{rk}(S) + \mbox{rk}(S')$.
\end{enumerate}
The independent sets are exactly those for which $\mbox{rk}(S)=|S|$ holds.
For the cycle matroid of a graph $G$ we can associate to a subset $S$ of $E$ the spanning sub-graph $H$
of $G$ obtained by taking all the vertices of $G$, but just the edges which are in $S$.
In this case the rank of $S$ equals the number of vertices of $H$ minus the number of connected components
of $H$.
The multivariate Tutte polynomial for a matroid is defined by
\bq
\label{def_Tutte_matroid}
\tilde{\mathcal Z}\left(q,x_1,...,x_n\right)
 & = &
 \sum\limits_{S \subseteq E} q^{-\mbox{rk}(S)}
 \prod\limits_{e_i \in S} x_i.
\eq
It is a polynomial in $1/q$ and $x_1$, ..., $x_n$.
Since a matroid can be defined by giving the rank for each subset $S$ of $E$, it is clear that
the multivariate Tutte polynomial encodes all information of a matroid.
For the cycle matroid of a graph $G$ with $r$ vertices the multivariate Tutte polynomial $\tilde{\cal Z}$
of the matroid is related to the multivariate Tutte polynomial ${\cal Z}$ of the graph by
\bq
\tilde{\mathcal Z}\left(q,x_1,...,x_n\right)
 & = &
 q^{-r} {\mathcal Z}\left(q,x_1,...,x_n\right).
\eq
For a matroid there are as well the notions of duality, deletion and contraction.
Let us start with the definition of the dual of a matroid. 
We consider a matroid $M$ with the ground set $E$ and the
set of bases $\mathcal{B}(M)=\{B_{1},\,,B_{2},\,...,\, B_{n}\}$.
The dual matroid $M^{\star}$ of $M$ is the matroid with ground set $E$ and whose set of
bases is given by $\mathcal{B}(M^{\star})=\{E-B_{1},\, E-B_{2},\,...,\, E-B_{n}\}.$
It should be noted that in contrast to graphs the dual matroid 
can be constructed for any arbitrary matroid.

Deletion and contraction for matroids are defined as follows:
Let us consider a matroid $M=(E,\,\mathcalI)$ with ground
set $E$ and $\mathcalI$ being the set of independent sets. 
Let us divide the ground set $E$ into two disjoint sets $X$ and $Y$:
\bq
 E = X \cup Y, & & X \cap Y = \emptyset.
\eq
We denote by ${\mathcalI}_Y$ the elements of ${\mathcalI}$, which are sub-sets of $Y$.
The matroid $M-X$ is then defined as the matroid with ground set $E-X=Y$ and whose set of independent sets
is given by ${\mathcalI}_Y$.
We say that the matroid $M-X$ is obtained from the matroid $M$ by deleting $X$.
The contracted matroid $M/X$ is defined as follows:
\bq
 M/X & = & \left( M^\ast - X \right)^\ast,
\eq
i.e. one deletes from the dual $M^\ast$ the set $X$ and takes then the dual.
The contracted matroid $M/X$ has the ground set $E-X$.
With these definitions of deletion and contraction we can now state the recursion relation for
the multivariate Tutte polynomial of a matroid:
We have to distinguish two cases. 
If the one-element set $\{e\}$ is of rank zero (corresponding to a self-loop in a graph) we have
\bq
\tilde{\mathcal Z}\left(M\right) 
 & = & \tilde{\mathcal Z}\left(M-\{e\}\right) + x_{e} \tilde{\mathcal Z}\left(M/\{e\}\right).
\eq
Otherwise we have 
\bq
\label{recursion_matroid_non_zero_rank}
\tilde{\mathcal Z}\left(M\right)
 & = & 
 \tilde{\mathcal Z}\left(M-\{e\}\right) + \frac{x_{e}}{q} \tilde{\mathcal Z}\left(M/\{e\}\right).
\eq
The recursion terminates for a matroid with an empty ground set, in this case we have $\tilde{\mathcal Z}=1$.
The fact that one has a different recursion relation for the case where the one-element set
$\{e\}$ is of rank zero is easily understood from the definition of $\tilde{\mathcal Z}$ and the
relation to graphs: 
For a cycle matroid $\tilde{\mathcal Z}$ differs from ${\mathcal Z}$ by the extra factor $q^{-r}$, where
$r$ is the number of vertices of the graph.
If $e$ is a self-loop of $G$, the contracted graph $G/e$ equals $G-e$ and in particular it has the same
number of vertices as $G$.
In all other cases the contracted graph $G/e$ has one vertex less than $G$, which is reflected by the 
factor $1/q$ in eq.~(\ref{recursion_matroid_non_zero_rank}).

\section{Conclusions}
\label{sect_conclusions}

In this review we considered polynomials associated to Feynman graphs. 
The integrand of a Feynman loop integral is directly related to two graph
polynomials, which are called the first and the second Symanzik polynomial.
Upon a simple substitution the first Symanzik polynomial is related to the
Kirchhoff polynomial of the graph.
The graph polynomials have many interesting properties. 
We discussed the interpretation in terms of spanning trees and spanning forests,
the all-minors matrix-tree theorem as well as recursion relations 
due to contraction and deletion of edges.
These properties in turn provide a variety of methods to compute the graph
polynomials.
We provided a detailed discussion of Dodgson's identity.
This identity provides a factorisation formula for a certain combination
of graph polynomials and can be used in practical calculations.
In addition we discussed the generalisation of the two graph polynomials 
towards the multivariate Tutte polynomial and matroids.
Matroid theory gives an answer to the question under which conditions two
inequivalent graphs have the same Kirchhoff polynomial.

\bibliography{/home/stefanw/notes/biblio}

\begin{thebibliography}{10}

\bibitem{Bloch:2005}
S.~Bloch, H.~Esnault, and D.~Kreimer,
\newblock Comm. Math. Phys. {\bf 267}, 181 (2006), math.AG/0510011.

\bibitem{Bloch:2008jk}
S.~Bloch and D.~Kreimer,
\newblock Commun. Num. Theor. Phys. {\bf 2}, 637 (2008), arXiv:0804.4399.

\bibitem{Bloch:2007}
S.~Bloch,
\newblock Japan J. Math. {\bf 2}, 165 (2007).

\bibitem{Bloch:2008}
S.~Bloch,
\newblock (2008), arXiv:0810.1313.

\bibitem{Andre:2008aa}
Y.~Andr{\'e},
\newblock (2008), arXiv:0812.3920.

\bibitem{Brown:2008}
F.~Brown,
\newblock Commun. Math. Phys. {\bf 287}, 925 (2008), arXiv:0804.1660.

\bibitem{Brown:2009a}
F.~Brown,
\newblock (2009), arXiv:0910.0114.

\bibitem{Brown:2009b}
F.~Brown and K.~Yeats,
\newblock (2009), arXiv:0910.5429.

\bibitem{Schnetz:2008mp}
O.~Schnetz,
\newblock (2008), arXiv:0801.2856.

\bibitem{Schnetz:2009}
O.~Schnetz,
\newblock (2009), arXiv:0909.0905.

\bibitem{Marcolli:2008cy}
M.~Marcolli and A.~Rej,
\newblock J. Phys. {\bf A41}, 315402 (2008), arXiv:0806.1681.

\bibitem{Aluffi:2008sy}
P.~Aluffi and M.~Marcolli,
\newblock Commun. Num. Theor. Phys. {\bf 3}, 1 (2009), arXiv:0807.1690.

\bibitem{Aluffi:2008rw}
P.~Aluffi and M.~Marcolli,
\newblock (2008), arXiv:0811.2514.

\bibitem{Aluffi:2009b}
P.~Aluffi and M.~Marcolli,
\newblock (2009), arXiv:0901.2107.

\bibitem{Aluffi:2009a}
P.~Aluffi and M.~Marcolli,
\newblock (2009), arXiv:0907.3225.

\bibitem{Marcolli_book}
M.~Marcolli,
\newblock {\em Feynman Motives} (World Scientific, 2010).

\bibitem{Bergbauer:2009yu}
C.~Bergbauer, R.~Brunetti, and D.~Kreimer,
\newblock (2009), arXiv:0908.0633.

\bibitem{Laporta:2002pg}
S.~Laporta,
\newblock Phys. Lett. {\bf B549}, 115 (2002), hep-ph/0210336.

\bibitem{Laporta:2004rb}
S.~Laporta and E.~Remiddi,
\newblock Nucl. Phys. {\bf B704}, 349 (2005), hep-ph/0406160.

\bibitem{Laporta:2008sx}
S.~Laporta,
\newblock Int. J. Mod. Phys. {\bf A23}, 5007 (2008), arXiv:0803.1007.

\bibitem{Bailey:2008ib}
D.~H. Bailey, J.~M. Borwein, D.~Broadhurst, and M.~L. Glasser,
\newblock (2008), arXiv:0801.0891.

\bibitem{Blumlein:2009}
J.~Bl{\"u}mlein, D.~J. Broadhurst, and J.~A.~M. Vermaseren,
\newblock Comput. Phys. Commun. {\bf 181}, 582 (2010), arXiv:0907.2557.

\bibitem{Belkale:2003}
P.~Belkale and P.~Brosnan,
\newblock Int. Math. Res. Not. , 2655 (2003).

\bibitem{Belkale:2003b}
P.~Belkale and P.~Brosnan,
\newblock Duke Math.~Journal {\bf 116}, 147 (2003).

\bibitem{Bierenbaum:2003ud}
I.~Bierenbaum and S.~Weinzierl,
\newblock Eur. Phys. J. {\bf C32}, 67 (2003), hep-ph/0308311.

\bibitem{Bogner:2007cr}
C.~Bogner and S.~Weinzierl,
\newblock Comput. Phys. Commun. {\bf 178}, 596 (2008), arXiv:0709.4092.

\bibitem{Bogner:2007mn}
C.~Bogner and S.~Weinzierl,
\newblock J. Math. Phys. {\bf 50}, 042302 (2009), arXiv:0711.4863.

\bibitem{Grozin:2008tp}
A.~G. Grozin,
\newblock (2008), arXiv:0809.4540.

\bibitem{Patterson:PhD}
E.~Patterson,
\newblock (2009), Ph.D. thesis, University of Chicago.

\bibitem{Kirchhoff:1847}
G.~Kirchhoff,
\newblock Ann. Physik Chemie {\bf 72}, 497 (1847).

\bibitem{Eden}
R.~J. Eden, P.~V. Landshoff, D.~I. Olive, and J.~C. Polkinghorne,
\newblock {\em The Analytic S-Matrix} (Cambridge University Press, 1966).

\bibitem{Hwa}
R.~C. Hwa and V.~L. Teplitz,
\newblock {\em Homology and Feynman Integrals} (W. A. Benjamin, 1966).

\bibitem{Nakanishi}
N.~Nakanishi,
\newblock {\em Graph Theory and Feynman Integrals} (Gordon and Breach, 1971).

\bibitem{Todorov}
I.~T. Todorov,
\newblock {\em Analytic Properties of Feynman Diagrams in Quantum Field Theory}
  (Pergamon Press, 1971).

\bibitem{Zavialov}
O.~I. Zavialov,
\newblock {\em Renormalized Quantum Field Theory} (Kluwer, 1990).

\bibitem{Smirnov:2006ry}
V.~A. Smirnov,
\newblock {\em Feynman integral calculus} (Springer, Berlin, 2006).

\bibitem{Itzykson:1980rh}
C.~Itzykson and J.~B. Zuber,
\newblock {\em Quantum Field Theory} (McGraw-Hill, New York, 1980).

\bibitem{Weinzierl:2006qs}
S.~Weinzierl,
\newblock Fields Inst. Commun. {\bf 50}, 345 (2006), hep-ph/0604068.

\bibitem{Tutte:1984}
W.~T. Tutte,
\newblock {\em Graph Theory}, Encyclopedia of mathematics and its applications
  Vol.~21 (Addison-Wesley, 1984).

\bibitem{Stanley:1998}
R.~P. Stanley,
\newblock Ann. Combin. {\bf 2}, 351 (1998).

\bibitem{Chaiken:1982}
S.~Chaiken,
\newblock SIAM J. Alg. Disc. Meth. {\bf 3}, 319 (1982).

\bibitem{Chen:1982}
W.~K. Chen,
\newblock {\em Applied graph theory, graphs and electrical networks} (North
  Holland, 1982).

\bibitem{Moon:1994}
J.~Moon,
\newblock Discrete Math. {\bf 124}, 163 (1994).

\bibitem{Godsil:2001}
C.~Godsil and G.~Royle,
\newblock {\em Algebraic Graph Theory} (Springer, 2001).

\bibitem{Sokal:2005aa}
A.~D. Sokal,
\newblock in: B. S. Webb, editor, Surveys in Combinatorics, Cambridge
  University Press  (2005), math/0503607.

\bibitem{Tutte:1947}
W.~T. Tutte,
\newblock Proc. Cambridge Phil. Soc. {\bf 43}, 26 (1947).

\bibitem{Tutte:1954}
W.~T. Tutte,
\newblock Can. J. Math. {\bf 6}, 80 (1954).

\bibitem{Tutte:1967}
W.~T. Tutte,
\newblock J. Combin. Theory {\bf 2}, 301 (1967).

\bibitem{Ellis-Monaghan:2008aa}
J.~Ellis-Monaghan and C.~Merino,
\newblock in: M. Dehmer, editor, Structural Analysis of Complex Networks,
  Birkh{\"a}user  (2010), arXiv:0803.3079.

\bibitem{Ellis-Monaghan:2008ab}
J.~Ellis-Monaghan and C.~Merino,
\newblock in: M. Dehmer, editor, Structural Analysis of Complex Networks,
  Birkh{\"a}user  (2010), arXiv:0806.4699.

\bibitem{Krajewski:2008aa}
T.~Krajewski, V.~Rivasseau, A.~Tanasa, and Z.~Wang,
\newblock (2008), arXiv:0811.0186.

\bibitem{Dodgson:1866}
C.~L. Dodgson,
\newblock Proc. Roy. Soc. London {\bf 15}, 150 (1866).

\bibitem{Zeilberger:1997}
D.~Zeilberger,
\newblock Electron. J. Combin. {\bf 4}, 2 (1997).

\bibitem{Stembridge:1998}
J.~Stembridge,
\newblock Ann. Combin. {\bf 2}, 365 (1998).

\bibitem{Kuratowski:1930}
C.~Kuratowski,
\newblock Fund. Math. {\bf 15}, 271 (1930).

\bibitem{Wagner:1937}
K.~Wagner,
\newblock Math. Ann. {\bf 114}, 570 (1937).

\bibitem{Diestel:2005}
R.~Diestel,
\newblock {\em Graph Theory} (Springer, 2005).

\bibitem{Oxley}
J.~Oxley,
\newblock {\em Matroid Theory} (Oxford University Press, 2006).

\bibitem{Oxley:2003aa}
J.~Oxley,
\newblock Cubo {\bf 5}, 179 (2003).

\bibitem{Whitney:1933aa}
H.~Whitney,
\newblock Amer. J. Math. {\bf 55}, 245 (1933).

\bibitem{Oxley:1986aa}
J.~Oxley,
\newblock in: N. White, editor, Theory of Matroids  (1986).

\bibitem{Truemper:1980aa}
K.~Truemper,
\newblock J. Graph Theory {\bf 4}, 43 (1980).

\end{thebibliography}
\bibliographystyle{/home/stefanw/latex-style/h-physrev5}

\end{document}